\documentclass[graybox,makeidx]{svmult}

\usepackage{mathptmx}       
\usepackage{helvet}         
\usepackage{courier}        
\usepackage{type1cm}        
\usepackage{makeidx}         
\usepackage{graphicx}        
\usepackage{multicol}        
\usepackage[bottom]{footmisc}

\usepackage{psfrag}

\makeindex             

\begin{document}
\title*{Flavor Superconductivity \& Superfluidity}
\author{Matthias Kaminski 
}
\institute{Matthias Kaminski \at Department of Physics, Princeton University, Princeton, NJ 08544, USA. \email{mkaminsk@princeton.edu}
}
\maketitle
\abstract{In these lecture notes we derive a generic holographic string theory realization of a p-wave superconductor and superfluid. 
For this purpose we also review basic D-brane physics, gauge/gravity methods at finite temperature, 
key concepts of superconductivity and recent progress in distinct 
realizations of holographic superconductors and superfluids. 
Then we focus on a D3/D7-brane construction yielding a superconducting 
or superfluid vector-condensate. The corresponding gauge theory is 3+1-dimensional
${\cal N}=2$ supersymmetric Yang-Mills theory with $SU(N_c)$ color and $SU(2)$ flavor symmetry. 
It shows a second order phase transition to a phase in which a $U(1)$ subgroup of the $SU(2)$ symmetry 
is spontaneously broken and typical superconductivity signatures emerge, such as 
a conductivity (pseudo-)gap and the Meissner-Ochsenfeld effect.
Condensates of this nature are comparable to those recently found experimentally in 
p-wave superconductors such as a \emph{ruthenate compound}. A string picture of the pairing mechanism and 
condensation is given using the exact knowledge of the corresponding field theory degrees of freedom.
}
\vfill
\section*{Contents}
\setcounter{minitocdepth}{3}
\dominitoc
\minitoc
\section{Introduction}
\label{kamsec:introduction}
In these lecture notes\footnote{These lecture notes are based on joint research work with Martin Ammon, 
Johanna Erdmenger, Patrick Kerner and Felix Rust.} we generically construct a holographic p-wave superconductor. This introductory
section serves to explain why this particular setup is of interest both from a string-theoretical 
as well as from a condensed matter physics point of view. In order to guide the unexperienced
reader in section \ref{kamsec:sch} we will describe the basic concepts in words and motivate the project.
Detailing this overview we build the full holographic setup from scratch in a self-contained manner in 
section \ref{kamsec:holoSetup}. Section \ref{kamsec:braneThermoSpec} briefly introduces common
holographic methods and then summarizes the major
results known for the thermodynamics and the spectrum of D-brane systems. With these outcomes
in the back of our minds we will understand how our system develops a new ground state
and why it can be interpreted as a $\rho$ meson superfluid or superconductor. We compute and
discuss thermodynamic observables, conductivities and the spectrum of our holographic superconductor in
section \ref{kamsec:signatures}. Finally, we put together all of our findings in order to draw
a string theory picture for the \emph{pairing mechanism} in our p-wave superconductor. 
These notes are self-contained and widely complementary to the notes focusing on holographic 
s-wave superconductors~\cite{kamHerzog:2009xv,kamHartnoll:2009sz,kamHorowitz:2010gk}.
%
\subsection{String motivation}
\label{kamsec:sMotivation}
The setup of intersecting D-branes and especially the particular D3-D7 construction presented here
is interesting because of its diverse applications. It has been successfully used to model strongly 
coupled particle physics phenomena such as a \emph{deconfinement}, or rather 
\index{meson melting}\emph{meson melting} phase transition for fundamental matter and
transport coefficients in the quark gluon plasma experimentally created at the RHIC collider in 
Brookhaven (see~\cite{kamKaminski:2008ai} for an introductory review and further references).
On the other hand it has recently been used to model strongly correlated electron systems
as they appear in superfluids and superconductors in the realm of condensed matter 
physics~\cite{kamAmmon:2008fc,kamAmmon:2009fe}.
All these applications are also crucial checks of the basic principles and methods coming from
the conjectured gauge/gravity correspondence. Receiving physically meaningful outcomes
when applying these holographic methods to various systems accessible by experiment, strengthens
our confidence in the gauge/gravity conjecture.

From the \emph{string-theoretic point of view} this particular setup is interesting because of its simplicity,
uniqueness (explained below) and the naturalness with which the symmetry is broken spontaneously here. 
The flavor\footnote{The term "flavor"
superconductor stems from earlier applications of this setup to model strongly correlated
high energy systems such as the quark gluon plasma. In the present case the name is not important
and possibly misguiding, since it is really only essential that the system has a non-Abelian $SU(2)$ symmetry.}  
superfluid/superconductor\footnote{We are using the terms superfluid and superconductor
interchangably here. For the considered phenomena this distinction does not make any difference.
Some details to this distinction are given in section \ref{kamsec:FTIdea}.} 
examined here was the first generic \emph{top-down} string theory realization 
of a superfluid/superconducting phase. In contrast to this the pioneering papers on holographic 
superconductors~\cite{kamHartnoll:2008vx,kamGubser:2008wv,kamHartnoll:2008kx} had
exclusively treated gravity toy models which were not directly obtained from string theory, i.\,e.\,
\index{bottom-up models}\emph{bottom-up models}. Therefore
the gauge/gravity correspondence could not be used to identify the exact gauge theory dual.
This fact obstructed the interpretation of the outcomes. Furthermore these toy models had few 
restrictions on their parameters such that in principle a large parameter space needed to be scanned,
see {e.\,g.\,}~\cite{kamHorowitz:2008bn}.
Our string-derived flavor superfluid/superconductor overcomes those two problems: First the
field theory degrees of freedom are exactly known since it is simply ${\cal{N}}=2$ supersymmetric
Yang-Mills theory with an $SU(2)$ flavor-symmetry. Second, the values of parameters, such
as for example the dimension of the condensing operator, in our setup are severly restricted 
by their string theoretic derivation. In this sense this setup is "unique" compared to the big parameter
space to be scanned in bottom-up toy models. Later, other top-down string realizations have been suggested and for
example involve a consistent truncation of type IIB supergravity with a 
chemical potential for the R-charge~\cite{kamGubser:2009qm}, 
and domain-wall solutions interpolating between AdS solutions with distinct radii
which may be lifted to IIB supergravity or eleven-dimensional supergravity~\cite{kamGubser:2009gp}.
%
\subsection{Condensed matter motivation}
\label{kamsec:cMotivation}
From a \emph{condensed-matter physics point of view} our flavor superconductor is highly interesting because it 
reproduces features which have been measured in experiments with unconventional superconductors,
such as p-wave superconductivity, a system of strongly-correlated particles, a pseudo-gap in the 
frequency-dependent conductivity~(found in high temperature d-wave superconductors). Most of 
these phenomena lack a widely-accepted microscopic explanation by conventional approaches, 
so there is the hope that gauge/gravity can shed some light on the nature of these systems. These systems
usually contain strongly correlated electrons, so the dual weak gravity description is in principle accessible.
Our system has a vector operator which condenses upon breaking a \emph{residual Abelian} flavor symmetry 
spontaneously. This gives the vector order parameter as described below. So there is a preferred spatial 
direction in the superfluid/superconducting condensate. This is exactly the situation recently found experimentally 
in the \emph{p-wave} (explained below in section \ref{kamsec:FTIdea})
superconductor $Sr_2 Ru O_4$~\cite{kamMaeno:1994na}. These materials are 
investigated with great excitement in the condensed-matter community because they 
are hoped to be usable for \index{quantum computing}\emph{quantum computing}~\cite{kamtewari-2007-98}. The reason is that
the p-wave structure implies the presence of non-Abelian quasi-particles in the \index{ruthenate compound}\emph{ruthenate compound}. These non-Abelian quasi-particles
can be used as the states to be manipulated in a \emph{topological quantum computer}. The biggest
practical obstacle for quantum computation are the errors which may occur during a calculation due
to materials being not ideal. Topological quantum computers minimize this source of error because
they carry out operations by \emph{braiding} the non-Abelian quasi-particles in a Hilbert-subspace 
containing degenerate ground states. Due to an energy gap between this subspace and the rest of the Hilbert space
it virtually decouples from all local perturbations~\cite{kamnayak-2008-80}. Another confirmed and well-studied
condensed-matter example for an emerging p-wave structure is superfluid ${}^3$He-A~\cite{kamVollhardt}.
%
\section{Superconductivity \& Holography}
\label{kamsec:sch}
This section is a primer on the subject of spontaneous symmetry breaking, superconductivity
and superfluidity in the holographic context of the gauge/gravity correspondence. Little formalism
is used, while we introduce all the necessary concepts.
%
\subsection{Basics of superconductivity \& our field theory idea}
\label{kamsec:FTIdea}
Let us review the essential concepts of superconductivity and understand how to build
a p-wave superconductor. We are going to need this knowledge in order to appreciate
the fact that our holographic setup reproduces this behavior in great detail.
\begin{table}
\caption{Nomenclature of superconducting states}
\label{kamtab:superCStates}   
\begin{tabular}{p{2cm}p{2.4cm}p{2cm}p{4.9cm}}
\hline\noalign{\smallskip}
Orbital angular momentum & Name & Parity of spatial part & Spin state  \\
\noalign{\smallskip}\svhline\noalign{\smallskip}
0 & s-wave & even & singlet\\
1 & p-wave & odd & triplet\\
2 & d-wave & even & singlet\\
\noalign{\smallskip}\hline\noalign{\smallskip}
\end{tabular}
\end{table}
\runinhead{Superconductor basics \& the p-wave} Superconductivity is the phenomenon associated
with infinite dc conductvity in materials at low temperatures. It is caused by the formation of a
\index{charged condensate}\emph{charged condensate} in which directed currents do not experience resistivity. A defining
criterion for superconductivity is the \index{Meissner-Ochsenfeld effect}\emph{Meissner-Ochsenfeld effect} described below.
In \index{conventional superconductors}\emph{conventional superconductors} the superconducting condensate consists of electron 
pairs called \emph{Cooper pairs}. So there are two simultaneous steps: the fermionic electrons have to pair up to
form bosonic Cooper pairs, and these pairs do condense. This condensation
happens in a second order phase transition (at vanishing magnetic field) when the temperature is lowered through its
critical value $T_c$. Due to the requirement
for the fermionic state to be antisymmetric there are only certain symmetry combinations 
allowed for the two electron state describing a Cooper pair.
As seen from table \ref{kamtab:superCStates} the name \index{p-wave}"\emph{p-wave}" superconductor refers to
those pairs in which the relative orbital angular momentum between the two electrons is 
$L=1$, the spatial part of the parity is odd and the spin state is a triplet. 

The mechanism pairing \index{pairing mechanism}electrons in conventional superconductors is well described by 
a mean field theory approach and the microscopic \index{BCS theory}\emph{BCS-theory}. Recall that condensed matter 
systems are conveniently described in terms of lattices with many (about $10^{23}$) sites.
Then conventional BCS-theory tells us that the lattice is slightly deformed by the presence of an electron.
This deformation can be described by a quasi-particle excitation, a \index{phonon}\emph{phonon}. We can
imagine the phonon to create a small potential well near the electron in which another electron
can be caught. So lattice vibrations (phonons) mediate a weakly attractive interaction between
the electrons which then form pairs. Conventional BCS-theory with phonons is only valid at low temperatures because around
$20 K$ the phononic lattice vibrations caused by the temperature already destroy the
weakly attractive interaction between the electrons. Note that BCS-theory itself does not depend on the origin 
of the attractive interaction.

\emph{A \index{superconductor}superconductor is simply a charged superfluid}. 
The crucial defining property for any kind of superconductivity or superfluidity is that a
symmetry is spontaneously broken. In superfluids this symmetry is global while in 
superconductors it is local, {i.\,e.\,} a gauge symmetry. Therefore in superconductors
there are a few additional effects related to the gauge symmetry and the corresponding
gauge field. But besides that those two phenomena are very similar. Especially in both 
cases there is a \index{Goldstone boson}\emph{Goldstone boson} created for each spontaneously broken
symmetry of the describing field theory. For a broken global symmetry this Goldstone
boson survives and is visible in the spectrum as a hydrodynamic mode~\cite{kamAmado:2009ts}. 
For a broken local symmetry however the Goldstone boson is eaten by the gauge
field which couples to the charge belonging to the broken symmetry. 
In superconductors this causes the gauge boson, {i.\,e.\,} the photon for the broken 
electromagnetic $U(1)$ to become massive. Since these heavy photons can travel 
only an exponentially small distance, the electromagnetic interaction becomes short-ranged.
Therefore magnetic fields, which can be thought of as consisting of photons, can only penetrate 
the system up to a certain distance, the \emph{penetration depth}. This is called
\emph{Meissner-Ochsenfeld effect} and it is a defining criterion for superconductivity. 
In the Anderson-Higgs mechanism particles acquire a mass by the same mechanism. Thus it is 
sometimes described as the superfluidity of the vacuum. See~\cite{kamgreiter-2005-319} for a
more precise review.

There is a class of experimentally well-studied but theoretically less understood 
\index{unconventional superconductors}\emph{unconventional superconductors}, such as copper or ruthenate compounds. Some of these
materials show superconducting phases at high temperatures\footnote{These 
\emph{high temperature superconductors} in general realize a \emph{d-wave} structure.
However, $Sr_2 Ru O_4$ belongs to a distinct class of unconventional superconductors and is
p-wave superconducting at low temperatures around $2K$. The pairing mechansim is not
microscopically understood. The simplified argument is that the phonon-interaction of
conventional Cooper pairing is isotropic, thus not providing an anisotropic p-wave structure.} up to $138 K$.
The conventional BCS-theory does not apply to such high temperatures as mentioned above. So the biggest mystery remains 
to understand the \index{pairing mechanism}\emph{pairing mechanism} of electrons in these high temperature
superconductors. Pairing and condensation need not occur at the same temperature here. 
Most important for the application of gauge/gravity duality: 
In these \emph{unconventional superconductors} the coupling strength of the electrons
to each other is strong. Thus these are experimentally accessible systems governed by strongly coupled field theory.

Another clear signature for superconductivity is of course an infinite dc conductivity. Together with
that there is a conductivity gap in the frequency-dependent conductivity. This is caused by 
the fact that the conductivity at small frequencies or energies vanishes until there is enough
energy to break up one Cooper pair. From that energy on the material is a normal conductor
with individual electrons being the charge carriers. In unconventional superconductors there
is a surprising phenomenon called \index{pseudo gap}\emph{pseudo-gap}. This means that the experiments
carried out in unconventional superconductors show a gap in the conductivity even
at and above the transition temperature where the superconducting condensate forms. 
Inside this pseudo-gap the conductivity does not drop to zero but to a small finite value.

\runinhead{How to build a field theory with p-wave superconductivity}
\begin{figure}[b]
\sidecaption
\includegraphics[scale=.65]{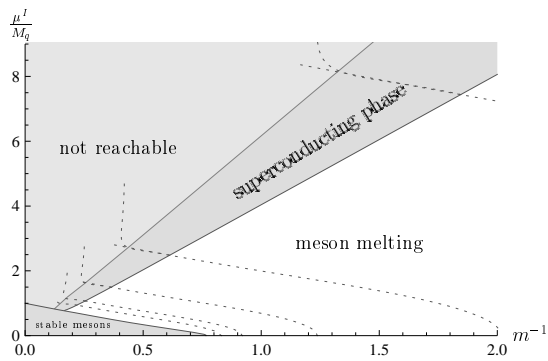}
\caption{Phase diagram of our field theory depending on the temperature parametrized
by $m^{-1}\propto T$ and on the isospin chemical potential $\mu^I$ scaled by the quark
mass $M_q$. At low temperatures and chemical potentials a phase of stable mesons forms
which melts at higher $\mu^I$ and $m^{-1}$. Above a critical density we find a superconducting
or superfluid phase with a vector order parameter.}
\label{kamfig:phasediagram1}
\end{figure}
As stressed above the crucial thing to do in order to get a superconductor or superfluid
is to spontaneously break a symmetry. We are going to accomplish this by allowing
our system to develop a charged condensate. 

Let us assume a particle physics point
of view for a while. In these notes we will focus on a 3+1 -- dimensional $\mathcal{N}=2$
supersymmetric $SU(N_c)$ Yang-Mills theory at temperature $T$, consisting of a
$\mathcal{N}=4$ gauge multiplet as well as $N_f$ massive $\mathcal{N}=2$
supersymmetric hypermultiplets $(\psi_i, \phi_i).$ The hypermultiplets give
rise to the \index{flavor}flavor degrees transforming in the fundamental representation of
the gauge group. The action is  written down explicitly for instance in
\cite{kamErdmenger:2007cm}. In particular, we work in the large $N_c$ limit with
$N_f \ll N_c$ at strong coupling, {i.\,e.\,} with $\lambda \gg 1,$ where $\lambda =
g_{\mathrm{YM}}^2 N_c$ is the 't Hooft coupling constant. In the following we will
consider only two flavors, {i.\,e.\,} $N_f=2.$ The flavor degrees of freedom are
called $u$ and $d$. If the masses of the two flavor degrees are degenerate,
the theory has a global $U(2)$ flavor symmetry, whose overall $U(1)_B$
subgroup can be identified with the baryon number. 

In the following we will consider the theory at finite isospin chemical potential $\mu$, which is introduced as the source of the operator
\begin{equation}
  \label{kameq:7}
  J^3_0\propto \bar\psi\sigma^3\gamma_0\psi+\phi\sigma^3\partial_0\phi=n_u-n_d\,,  
\end{equation}
where $n_{u/d}$ is the charge density of the isospin fields,
$(\phi_u,\phi_d)=\phi$ and $(\psi_u,\psi_d)=\psi$. $\sigma^i$ are the Pauli matrices. A non-zero vev $\langle J^3_0\rangle$ introduces an isospin density as discussed in
\cite{kamErdmenger:2008yj}. The isospin chemical potential $\mu$ explicitly breaks the $U(2)\simeq
U(1)_B\times SU(2)_I$ flavor symmetry down to $U(1)_B\times U(1)_3$, where
$U(1)_3$ is generated by the unbroken generator $\sigma^3$ of the $SU(2)_I$.
Under the $U(1)_3$ symmetry the fields with index $u$ and $d$ have positive
and negative charge, respectively. 

However, the theory is unstable at large isospin chemical potential
\cite{kamErdmenger:2008yj}. The new phase is sketched in figure
\ref{kamfig:phasediagram1}. We show in this lecture (see also \cite{kamAmmon:2008fc}),
that the new phase is stabilized by a non--vanishing vacuum expectation value
of the current component 
\begin{equation}
  \label{kameq:1}
    J^1_3\propto \bar\psi\sigma^1\gamma_3\psi+\phi\sigma^1\partial_3\phi=\bar\psi_u\gamma_3 \psi_d+\bar \psi_d\gamma_3 \psi_u+\mathrm{bosons} \, .
 \end{equation}
This current component breaks both the $SO(3)$ rotational symmetry as well as 
the remaining Abelian $U(1)_3$ flavor symmetry spontaneously. The rotational $SO(3)$ is broken down 
to $SO(2)_3$, which is generated by rotations around the $x^3$ axis. Due to
the non--vanishing vev for $J^1_3,$ flavor charged vector mesons condense and
form a superfluid. Let us emphasize that we do not describe a color
superconductor on the field theory side, since the condensate is a gauge
singlet. Figure~\ref{kamfig:phasediagram1} shows a "not accessable" parameter region in which 
we get divergent quantities. This is due to the fact that at such large charge densities we would need
to take in account the backreaction of the D$7$-branes on the AdS geometry.

In a condensed matter context our model can be considered as a 
holographic p--wave superconductor in the following way.
The global $U(1)_3$ in our model is the analog of the local $U(1)_{\mathrm{em}}$
symmetry of electromagnetic interactions. So far in all holographic models of superconductors
the breaking of a global symmetry on the field theory side is considered. In our 
model, the current $J^3$ corresponds to the electric 
current $J_{\mathrm{em}}.$ The condensate $\langle J^1_3\rangle$ breaks the $U(1)_3$ 
spontaneously. Therefore it can be viewed as the superconducting condensate, 
which is analogous to the Cooper pairs. Since the 
condensate $\langle J^1_3\rangle$ transforms as a vector under spatial 
rotations, it is a p--wave superconductor.
-- Strictly speaking, for a \index{superconductor}superconductor interpretation it would be necessary to charge
the \index{superfluid}superfluid, i.\,e.\, gauge the global $U(1)_3$ symmetry which is broken spontaneously in
our model. A spontaneously broken global symmetry corresponds to a
superfluid.   However, as we mentioned before many features of 
superconductivity do not depend on whether the 
$U(1)_3$ is gauged. One exception to this is the Meissner--Ochsenfeld effect. 
To generate the currents expelling the magnetic field, the $U(1)_3$
symmetry has to be gauged. This matter is discussed further below in section~\ref{kamsec:meissnerO}
where we will be able to see the onset of the Meissner--Ochsenfeld effect within our holographic setup. 
%
\subsection{Holographic realization}
\label{kamsec:holoReal}
Holographic superconductors have first been studied in~\cite{kamGubser:2008px,kamGubser:2008zu,kamHartnoll:2008vx,kamGubser:2008wv}.
The initial idea presented in~\cite{kamGubser:2008px} was that the Abelian Higgs model coupled to gravity with a negative cosmological
constant provides a charged scalar condensate near but outside a charged black hole horizon. The charged condensate
spontaneously breaks the Abelian gauge symmetry of the theory. Later studies revealed that this breaking also
occurs in setups with neutral black holes but with a negative mass for the charged scalar~\cite{kamHartnoll:2008vx}.
The basic idea for a holographic p-wave superconductor has been outlined in~\cite{kamGubser:2008wv}.
Let us review the basic ideas.

\runinhead{Holographic superconductor basics}
In order to spontaneously break a symmetry we need a gravity theory with a gauge symmetry. 
Note that this is not the usual gauge symmetry $SU(N)$ of the correspondence where $N\to\infty$,
but an additional one with a finite rank, for example an Abelian $U(1)$.
Furthermore we need a black hole background in order to introduce finite temperature in the
dual field theory, see~\cite{kamKaminski:2008ai} for details. Most importantly we need a charged 
condensate hovering in the bulk over the horizon. To be more precise, we need a bulk field $\phi$ charged under
the gauge symmetry which is to be broken. $\phi$ has to have the typical expansion 
$\phi=\phi_{\mathrm{normalizable}}+\phi_{non-normalizable}+\dots$
near the AdS boundary, but with $\phi_{\mathrm{non-normalizable}}\equiv 0$. Why do we require this 
particular structure? Recall that in the gauge/gravity
correspondence the normalizable mode is identified with the field theory \emph{expectation value} $\langle {\cal O_\phi}\rangle$ 
of the operator ${\cal O_{\phi}}$ dual to the gravity field $\phi$. This is our condensate in the field theory,
which we want to be non-zero. The normalizable mode on the other hand is dual to a \emph{source} in the field 
theory. Therefore we want it to vanish since it would break the symmetry \emph{explicitly}. The role of
the field $\phi$ providing the condensate could be played for example by a charged scalar or by
one component of a gauge field. These concepts are illustrated in the following example.

\emph{\runinhead{Example:} Consider the bottom-up toy model p-wave superconductor introduced in~\cite{kamGubser:2008wv}.
There we have an Einstein Yang-Mills theory with a negative cosmological constant
\begin{equation}
\label{kameq:EYMaction}
S\sim\int \D^4 x \left [ 
R-\frac{1}{4}(F_{\mu\nu}^a)^2 +\frac{6}{L^2},
\right ]
\end{equation}
with $F$ being the field strength of an $SU(2)$ gauge field $A$. This $SU(2)$ in this case is the gauge 
group that we want to break. Actually we break only a $U(1)$ subgroup of it. But let us not worry
about the details now. Only note that this theory is placed in an AdS${}_4$ charged black hole background.
Our gauge field $A$ now plays the role of the field $\phi$ providing the condensate.
The operator dual to the gauge field $A$ is the electromagnetic current ${\cal O_\phi}=J^\mu$. 
The boundary behavior of the gravity field's components according to the equations of motion
derived from equation~(\ref{kameq:EYMaction}) is given by
\begin{eqnarray}
A^3_t= \mu+ \frac{d}{r} \, ,\quad A^1_x=0 + \frac{\langle J^1_x \rangle}{r}\, ,
\end{eqnarray}
with the radial AdS coordinate $r$. In principle we could allow more non-vanishing components
but this combination turns out to be both sufficient and consistent.
As usual in thermal AdS/CFT the temporal component $A_t$ introduces a chemical potential $\mu$
in the dual field theory. This chemical potential sources the corresponding operator $J^t=d$ which is
simply the charge density of the $SU(2)$ charge in the field theory at the boundary. This explicitly
breaks $SU(2)\to U(1)_3$. However the spatial component $A^1_x$ spontaneously breaks 
the remaining $U(1)_3$ with the condensate $\langle J^1_x \rangle$.
}

\runinhead{How to build a gravity dual to p-wave superconductivity}
As mentioned several times before, we need a vector condensate for our p-wave superconductor.
Conveniently the previous example already showed us what the structure for the dual gravity
theory has to be in order to realize a vector condensate. Unfortunately that example has not been derived from string theory.
Probably the most difficult task in building a holographic superconductor is finding a gravity
setup with all the neccessary features, which is actually stable and thermodynamically favored.
We are going to see that the system of intersecting D3 and D7 branes provides exactly
such a stable configuration. Since we are going to review Dp/Dq- brane systems below,
let us here spot only those features which are of importance to superconductivity. The $SU(2)$ 
"flavor" gauge group which we are going to break (partly spontaneously) is created by using two coincident
D7 branes. By introducing a gauge field $A^a_\mu$ living on these D7-branes and giving its 
temporal component a non-trivial profile in one of the flavor directions, 
we introduce a chemical potential in the dual field theory. This is
completely analogous to introducing the non-trivial $A_t^3$ in the previous example.
All this is going to take place in the background of a stack of $N_c$ black D3-branes. They 
introduce the temperature into the dual field theory. Finally, the gravity field which will 
break the residual flavor symmetry is going to be a spatial component of the 
gauge field just as $A_x^1$  in the previous example. Our analysis of the thermodynamic
potentials is going to show that this phase is thermodynamically preferred over the 
phase without the symmetry-breaking condensate. Furthermore it is stable against all 
obvious gauge field fluctuations. 
%
\section{Holographic setup}
\label{kamsec:holoSetup}
In this section we carry out exactly the D3-D7 brane construction outlined in section \ref{kamsec:holoReal}. The
result will be a gravity setup being holographically dual to a p-wave superfluid/superconductor. But
let us first review how to add flavor to the gauge/gravity correspondence.
%
\subsection{Flavor from intersecting branes}
\label{kamsec:flavor}
Let us imagine for this subsection that we want to use gauge/gravity in order to model QCD or 
the quark gluon plasma state of matter produced at the RHIC Brookhaven heavy-ion collider. 
The original AdS/CFT conjecture does not include matter in the fundamental representation of the gauge 
group but only adjoint matter. In order to come closer to a QCD-like behavior one can therefore investigate how
to incorporate quarks and their bound states in this section. We focus on the main results of~\cite{kamKarch:2002sh}
and~\cite{kamKruczenski:2003be}, however for a concise review the reader is referred to~\cite{kamErdmenger:2007cm}.

Since AdS/CFT has been discovered a lot of modifications of the original conjecture have 
been proposed and analyzed. This is always achieved by modifying the gravity theory in an appropriate way. 
For example the metric on which the gravity theory is defined may be changed to produce chiral symmetry breaking
in the dual gauge theory~\cite{kamConstable:1999ch,kamBabington:2003vm}. Other modifications put the 
gauge theory at finite temperature and produce 
confinement~\cite{kamWitten:1998zw}. Besides the introduction of finite temperature the inclusion of fundamental
matter, i.e. quarks, is the most relevant extension for us since we are aiming at a qualitative description
of strongly coupled QCD effects at finite temperature. These effects are similar to the ones observed at the
RHIC~heavy ion collider.

\runinhead{Adding flavor to AdS/CFT}
The change we have to make on the gravity side in order to produce fundamental matter on the gauge
theory side is the introduction of a small number~$N_f$ of D7-branes. These are also called \index{probe branes}\emph{probe
branes} since their backreaction on the geometry originally produced by the stack of $N$ D3-branes
is neglected. Strings within this D3/D7-setup now have the choice of starting~(ending) on the
D3- or alternatively on the D7-brane.
Note that the two types of branes share the four Minkowski directions~$0,1,2,3$ in which also the dual gauge theory
will extend on the boundary of AdS as visualized in figure~\ref{kamfig:d3D7Coords}.
\begin{figure}[b]
\begin{center}
\begin{tabular}[c]{|c|c|c|c|c|c|c|c|c|c|c|}
\hline
 & 0 & 1 & 2 & 3 & 4 & 5 & 6 & 7 & 8 & 9 \\
 \hline
 D3 & x & x & x & x  & & & & & & \\
 \hline
 D7 & x & x & x & x & x & x & x & x&  & \\
 \hline
\end{tabular} 
\end{center}
 \caption{ \label{kamfig:d3D7Coords}
 Coordinate directions in which the D$p$-branes extend are marked by 'x'. D3- and D7-branes 
 always share the four Minkowski directions and may be separated in the $8,9$-directions 
 which are orthogonal to both brane types.
 }
\end{figure}
The configuration of one string ending on $N$ coincident D3-branes produces an $SU(N)$~gauge symmetry of 
rotations in color space. Similarly the $N_f$ D7-branes generate a $U(N_f)$~flavor gauge
symmetry. We will call the strings starting on the stack of D$p$-branes 
and ending on the stack of D$q$-branes $p-q$ strings. The original $3-3$ strings are unchanged while
the $3-7$- or equivalently $7-3$ strings are interpreted as quarks on the gauge theory side of the correspondence.
This can be understood by looking at the $3-3$ strings again. They come in the adjoint representation of the
gauge group which can be interpreted as the decomposition of a bifundamental 
representation~$(N^2 -1)\oplus 1 = N \otimes \bar N$. So the two string ends on the D3-brane are interpreted
as one giving the fundamental, the other giving the anti-fundamental representation in the gauge theory.
In contrast to this the $3-7$ string has only one end on the D3-brane stack corresponding to a single 
fundamental representation which we interpret as a single quark in the gauge theory. 

We can also give mass to these quarks by separating the stack of D3-branes from the D7-branes
in a direction orthogonal to both branes. Now $3-7$ strings are forced to have a finite length~$L$
which is the minimum distance between the two brane stacks. On the other hand a string is an
object with tension and if it assumes a minimum length, it needs to have a minimum energy being
the product of its length and tension. The dual gauge theory object is the quark and it now also 
has a minimum energy which we interpret as its mass~$M_q= L/(2\pi \alpha')$.

The $7-7$ strings decouple from the rest of the theory since their effective coupling is suppressed 
by~$N_f/N$. In the dual gauge theory this limit corresponds to neglecting quark loops which is often 
called \emph{the quenched approximation}. Nevertheless, they are important for the description of 
mesons as we will see below.

Let us be a bit more precise about the fundamental matter introduced by $3-7$ strings. The gauge
theory introduced by these strings (in addition to the original setup) gives a ${\cal N}=2$ supersymmetric
$U(N)$ gauge theory containing~$N_f$ fundamental hypermultiplets.

\runinhead{D7 embeddings \& meson excitations}
Mesons correspond to fluctuations of the D7-branes\footnote{To be precise the fluctuations 
correspond to the mesons with spins 0, 1/2 and 1~\cite{kamKruczenski:2003be,kamKirsch:2006he}.} embedded 
in the $AdS_5\times S^5$-background generated by the D3-branes. 
From the string-point of view these fluctuations are
fluctuations of the hypersurface on which the $7-7$ strings can end, hence these are small 
oscillations of the $7-7$ string ends. The $7-7$ strings again lie in the adjoint representation of
the flavor gauge group for the same reason which we employed above to argue that $3-3$ strings
are in the adjoint of the (color) gauge group. Mesons are the natural objects in the adjoint flavor representation.
Vector mesons correspond to fluctuations of the gauge field on the D7-branes.

Before we can examine mesons as D7-fluctuations we need to find out how the D7-branes are 
embedded into the 10-dimensional geometry without any fluctuations. Such a stable configuration needs to
minimize the effective action. The effective action to consider is the world volume action of the D7-branes
which is composed of a Dirac-Born-Infeld 
and a topological Chern-Simons part
\begin{eqnarray}
\label{kameq:d7MesonAction}
S_{\mathrm{D}7}=& -T_{\mathrm{D}7} \int \D^{8}\sigma e^{-\Phi}
 \sqrt{
  -\det  \left \{ 
  P[g+B]_{\alpha\beta} + 
  (2\pi \alpha') F_{\alpha\beta}  
\right \}
 }\\ 
&  + \frac{(2\pi \alpha')^2}{2} T_{\mathrm{D7}} \int P[C_4]\wedge F \wedge F\, .
\end{eqnarray}
The preferred coordinates to examine the fluctuations of the D7 are obtained from the standard
AdS coordinates
\begin{equation}
\label{kameq:standardMetric}
\D s^2 = \frac{R^2}{\varrho^2}\D \varrho^2 + \frac{\varrho^2}{R^2}(-\D t^2 + {\D \vec x}^2)\, ,
\end{equation}
with the AdS radius $R$ and the dimensionful radial AdS coordinate $\varrho$.
\runinhead{Exercise}
\emph{Show that the standard AdS metric~(\ref{kameq:standardMetric}) transforms to 
\begin{equation}
\label{kameq:metricWithW}
{\D s}^2= \frac{r^2}{R^2}{\D \vec x}^2 + \frac{R^2}{r^2} ({\D \varrho}^2 + \varrho^2 {\D \Omega_3}^2
 + {\D w_5}^2 + {\D w_6}^2) \, , 
\end{equation}
under the transformation~$\varrho^2 = {w_1}^2 + \dots + {w_4}^2 , \,
r^2=\varrho^2+{w_5}^2+{w_6}^2$,
where~$\vec x$ is a four vector in Minkowski directions~$0,1,2,3$ and $R$ is the AdS radius.
The coordinate~$r$ is the radial AdS coordinate while~$\varrho$ is the radial coordinate on the 
coincident D7-branes.
}

Let us follow~\cite{kamKruczenski:2003be}: For a static D7 embedding with vanishing field strength~$F$ on the D7 world volume
the equations of motion are
\begin{equation}
\label{kameq:w56EOM}
0 = \frac{\D}{\D\varrho} \left ( \frac{\varrho^3}{\sqrt{1+{w_5'}^2+{w_6'}^2}}\frac{\D w_{5,6}}{\D \varrho} \right ) \, ,
\end{equation}
where~$w_{5,6}$ denotes that these are two equations for the two possible directions of fluctuation.
Since~(\ref{kameq:w56EOM}) is the equation of motion of a supergravity field in the bulk, 
the solution near the AdS boundary takes the standard form with a non-normalizable and a normalizable mode,
or source and expectation value respectively
\begin{equation}
w_{5,6} = m + \frac{c}{\varrho^2} +\dots \, ,
\end{equation}
with $m$~being the quark mass acting as a source and $c$~being the expectation value of the operator 
which is dual to the field~$w_{5,6}$. While $c$ can be related to the scaled quark 
condensate~$c\propto \langle\bar q q\rangle(2\pi\alpha')^3$.

If we now separate the D7-branes from the stack of D3-branes the quarks become massive and the 
radius of the~$S^3$ on which the D7 is wrapped becomes a function of the radial AdS coordinate~$r$.
The separation of stacks by a distance~$m$ modifies the metric induced on the D7~$P[g]$ such that it contains  
the term~$R^2 \varrho^2/(\varrho^2+m^2){\D\Omega_3}^2$. This expression vanishes at a 
radius~$\varrho^2=r^2-m^2=0$ such that the~$S^3$ shrinks to zero size at a finite AdS radius. 

Fluctuations about these~$w_{5}$ and~$w_6$ embeddings give scalar and pseudoscalar mesons.
We take
\begin{equation}
w_5 = 0 + 2\pi\alpha' \chi\, , \quad w_6 = m + 2\pi\alpha' \varphi
\end{equation}
After plugging these into the effective action~(\ref{kameq:d7MesonAction}) and expanding to quadratic
order in fluctuations we can derive the equations of motion for~$\varphi$ and~$\chi$. As an example 
we consider scalar fluctuations using an Ansatz
\begin{equation}
\label{kameq:mesonAnsatz}
\varphi = \phi (\varrho) e^{i \vec k\cdot \vec x} \mathcal{Y}_l(S^3)\, ,
\end{equation}
where~$\mathcal{Y}_l(S^3)$ are the scalar spherical harmonics on the~$S^3$, $\phi$ solves the radial
part of the equation and the exponential represents propagating waves with real momentum~$\vec k$.
We additionally have to assume that the mass-shell condition 
\begin{equation}
M^2 = -{\vec k}^2
\end{equation}
is valid. Solving the radial part of the equation we get the 
hypergeometric function~$\phi \propto F(-\alpha,\, -\alpha+l+1;\, (l+2);\, \frac{-\varrho^2}{m^2})$ and
the parameter
\begin{equation}
\alpha=-\frac{1-\sqrt{1-{\vec k}^2 R^4/m^2}}{2}
\end{equation}
summarizes a factor 
appearing in the equation of motion. In general this
hypergeometric function may diverge if we take~$\varrho\to \infty$. But since this is not compatible with our 
linearization of the equation of motion in small fluctuations, we further demand normalizability of the 
solution. This restricts the sum of parameters appearing in the hypergeometric function to take the integer values
\begin{equation}
n = \alpha - l -1  \, , \quad n=0,\, 1,\, 2, \dots \, .
\end{equation}
With this quantization condition we determine the scalar meson mass spectrum to be
\begin{equation}
\label{kameq:Ms}
M_{\mathrm{(pseudo)scalar}} =  \frac{2 m}{R^2} \sqrt{(n+l+1)(n+l+2)}\, ,
\end{equation}
where~$n$ is the radial excitation number found for the hypergeometric function.
Similarly we can determine pseudoscalar masses and find the same formula~\ref{kameq:Ms}.
For vector meson masses we need to consider fluctuations of the gauge field~$A$ appearing in
the field strength~$F$ in equation~(\ref{kameq:d7MesonAction}). The formula for vector mesons
(corresponding to e.g. the $\varrho$-meson of QCD) is 
\begin{equation}
\label{kameq:susyVectorMesonM}
M_{\mathrm{v}} = \frac{2 m}{R^2}\sqrt{(n+l+1)(n+l+2)} \, .
\end{equation}
Note that the scalar, pseudoscalar and vector mesons computed within this framework show 
identical mass spectra. Further fluctuations corresponding to other mesonic excitations can be 
found in~\cite{kamKruczenski:2003be,kamKirsch:2006he}.

\runinhead{Brane embeddings at finite temperature}
In order to get a finite temperature in the dual field theory, the gravity theory needs to be put into
a black hole or black brane background geometry.
It was found in~\cite{kamBabington:2003vm,kamMateos:2006nu,kamKirsch:2004km} that at finite 
temperature our $N_f$ probe flavor branes can be embedded 
in two distinct ways. There are high temperature configurations called \index{black hole embeddings}\emph{black hole embeddings} 
in which part of the brane falls into the black hole horizon. On the other hand there are low-temperature
configurations called \index{Minkowski embeddings}\emph{Minkowski embeddings} in which the brane stays outside the black hole
horizon. These two configurations are separated by a geometric transition, {i.\,e.\,} the configuration
in which the brane just barely touches the black hole horizon. This geometric transition corresponds
to the meson melting transition for the fundamental matter of the dual field theory. Note that the
adjoint matter of this field theory is always deconfined in this setup. 

At finite charge density however there is only one kind of embedding and that is the black hole embedding.
The heuristic argument is that introducing a finite charge on the brane there have to be field lines 
for the associated field strength. These lines have to end somewhere. Our setup has rotational 
symmetry in the spatial directions. Imagine the radial AdS coordinate and field lines running along it. 
If they are supposed to end somewhere, there has to be a horizon. Otherwise they would all 
meet in the origin at $\rho=0$. Note that field lines ending at the horizon means, that we can 
interpret the horizon as being charged. In these lecture notes we will exclusively deal with
non-vanishing density and thus only encounter black hole embeddings.
%
\subsection{Background and brane configuration} 
\label{kamsec:backgr-brane-conf}
We aim to have a field theory dual at finite temperature. This is holographically accomplished
by placing the gravity theory in a black hole or black brane background. Here the black hole's
Hawking temperature can be identified with the field theory temperature.  
We consider asymptotically $AdS_5\times S^5$ space-time. The $AdS_5\times
S^5$ geometry is holographically dual to the ${\cal N}=4$ Super Yang-Mills
theory with gauge group $SU(N_c)$. The dual description of a finite
temperature field theory is an AdS black hole. We use the coordinates of
\cite{kamKobayashi:2006sb} to write the AdS black hole background in Minkowski
signature as 
\begin{equation}
  \label{kameq:AdSmetric}
  \D s^2=\frac{\varrho^2}{2R^2}\left(-\frac{f^2}{\tilde{f}}\mathrm{d}
    t^2+\tilde{f}
    \mathrm{d}\vec{x}^2\right)+\left(\frac{R}{\varrho}\right)^2(\mathrm{d}\varrho^2+\varrho^2\mathrm{d}\Omega_5^2)\,,
\end{equation}
with $\mathrm{d}\Omega_5^2$ the metric of the unit 5-sphere and
\begin{equation}
  f(\varrho)=1-\frac{\varrho_H^4}{\varrho^4},\quad
  \tilde{f}(\varrho)=1+\frac{\varrho_H^4}{\varrho^4}\,, 
\end{equation}
where $R$ is the AdS radius, with
$R^4=4\pi g_s N_c\,{\alpha'}^2 = 2\lambda\,{\alpha'}^2$.
\emph{
\runinhead{Exercise:}
Show that the metric~(\ref{kameq:AdSmetric}) can be obtained from the standard finite
temperature AdS metric
\begin{equation}
\D s^2=\frac{(\pi T R)^2}{u} [-f(u){\D t}^2 +{\D \mathbf{x}}^2] 
 + \frac{R^2}{4 u^2 f(u)}{\D u}^2+R^2 {\D\Omega_5}^2 \, ;\quad f(u)=1-u^2\, ,
\end{equation}
by the transformation $(r_0\rho)^2=2 r^2 + \sqrt{r^4-{r_0}^4}$, with $r$ being the original 
radial AdS coordinate and $r_0$ the location of the black hole horizon.
}

\emph{ \runinhead{Exercise:}
The temperature of the black hole given by (\ref{kameq:AdSmetric}) may be
determined by demanding regularity of the Euclidean section. Show that it is given by
\begin{equation}
  T=\frac{\varrho_H}{\pi R^2}\,. 
\end{equation}}
 
In the following we may use the dimensionless coordinate
$\rho=\varrho/\varrho_H$, which covers the range from the event horizon at
$\rho=1$ to the boundary of the AdS space at $\rho\to\infty$.
\begin{figure}[b]
\sidecaption
       \includegraphics[width=0.6\textwidth]{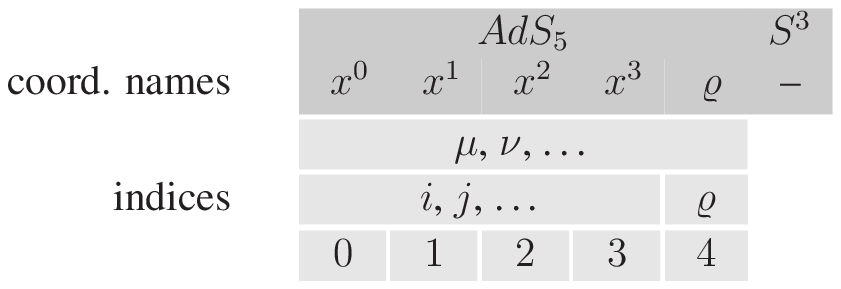}
       \caption{This figure summarizes our coordinates and indices.}
\end{figure}
To include fundamental matter, we embed $N_f$ coinciding D$7$-branes into
the ten-dimensional space-time as illustrated in figure \ref{kamfig:d3D7Coords}.
These D$7$-branes host flavor gauge fields
$A_\mu$ with gauge group $U(N_f)$. To write down the DBI action for the
D$7$-branes, we introduce spherical coordinates $\{r,\Omega_3\}$ in the
4567-directions and polar coordinates $\{L,\phi\}$ in the 89-directions
\cite{kamKobayashi:2006sb}. The angle between these two spaces is denoted by
$\theta$ ($0\le\theta\le\pi/2$). The six-dimensional space in the
$456789$-directions is given by
\begin{equation}
    \mathrm{d}\varrho^2+\varrho^2\mathrm{d}\Omega_5^2=\,\mathrm{d} r^2+r^2\mathrm{d}\Omega_3^2+\mathrm{d} L^2+L^2\mathrm{d}\phi^2=\,\mathrm{d}\varrho^2+\varrho^2(\mathrm{d}\theta^2+\cos^2\theta\mathrm{d}\phi^2+\sin^2\theta\mathrm{d}\Omega_3^2)\,,
\end{equation}
where $r=\varrho\sin\theta$, $\varrho^2=r^2+L^2$ and $L=\varrho\cos\theta$. D3- and D7-branes 
always share the four Minkowski directions and may be separated in the $8,9$-directions 
which are orthogonal to both brane types. That separation is dual to the mass of the 
{fundamental flavor fields} in the dual gauge theory, i.\,e.\, the {\index{quarks}\emph{quarks}}.

Due to the $SO(4)$ rotational symmetry in the 4567 directions, the embedding of the D$7$-branes only depends on the
radial coordinate $\rho$. Defining $\chi=\cos\theta$, we parametrize the
embedding by $\chi=\chi(\rho)$ and choose $\phi=0$ using the $SO(2)$
symmetry in the 89-direction. The induced metric $G$ on the D$7$-brane
probes is then
\begin{equation}
  \label{kameq:inducedmetric}
    ds^2(G)=\frac{\varrho^2}{2R^2}\left(-\frac{f^2}{\tilde{f}}\mathrm{d}
      t^2+\tilde{f}\mathrm{d}\vec{x}^2\right)+\frac{R^2}{\varrho^2}\frac{1-\chi^2+\varrho^2(\partial_\varrho\chi)^2}{1-\chi^2}\mathrm{d}\varrho^2+R^2(1-\chi^2)\mathrm{d}\Omega_3^2\,.
\end{equation}
The square root of the determinant of $G$ is given by
\begin{equation}
  \sqrt{-G}=\frac{\sqrt{h_3}}{4}\varrho^3f\tilde{f}(1-\chi^2)\sqrt{1-\chi^2+\varrho^2(\partial_\varrho\chi)^2}\,,
\end{equation}
where $h_3$ is the determinant of the 3-sphere metric.

As in \cite{kamErdmenger:2008yj} we introduce a $SU(2)$ isospin chemical
potential $\mu$ by a non-vanishing time component of the non-Abelian background field
on the D$7$-brane. The generators of the $SU(2)$ gauge group are given by the
Pauli matrices $\sigma^i$. Due to the gauge symmetry, we may rotate the flavor
coordinates until the chemical potential lies in the third flavor direction,
\begin{equation}
  \label{kameq:isospinchempot}
  \mu=\lim_{\rho\to\infty}A_0^3(\rho)\,.
\end{equation}
This non-zero gauge field breaks the $SU(2)$ gauge symmetry down to $U(1)_3$
generated by the third Pauli matrix $\sigma^3$. The spacetime symmetry on the
boundary is still $SO(3)$. Notice that the Lorentz group $SO(3,1)$ is already broken down
to $SO(3)$ by the finite temperature. In addition, we consider a further
non-vanishing background gauge field which stabilizes the system for large
chemical potentials. Due to the symmetry of our setup we may choose $A_3^1\mathrm{d}
x^3\sigma^1$ to be non-zero. To obtain an isotropic configuration in the field
theory, this new gauge field $A_3^1$ only depends on $\rho$. Due to this two
non-vanishing gauge fields, the field strength tensor on the branes has the
following non-zero components,
\begin{svgraybox}
\begin{eqnarray}
  \label{kameq:nonzeroF}
    &F^1_{\varrho 3}=-F^1_{3\varrho}=\partial_\varrho A^1_3\qquad \mathrm{(7-7 strings)}\,,\\
    &F^3_{\varrho 0}=-F^3_{0\varrho}=\partial_\varrho A^3_0\qquad \mathrm{(3-7 strings)}\,,\\
    &F^2_{03}=-F^2_{30}=\frac{\gamma}{\sqrt{\lambda}}A^3_0A^1_3\qquad \mathrm{(interaction)}\,.
\end{eqnarray}
\end{svgraybox}
The labels behind those equations refer to the sort of strings which generate the
corresponding gauge fields. The field strength $F_{03}^2$ can be understood as an interaction
term between 7--7 and 3--7 strings. We derive this interpretation in section~\ref{kamsec:string-theory-pict}.
%
\subsection{DBI action and equations of motion}
\label{kamsec:dbi-action-equations}
In this section we calculate the equations of motion which determine the
profile of the D$7$-brane probes and of the gauge fields on these branes. A
discussion of the gauge field profiles, which we use to give a geometrical
interpretation of the stabilization of the system and the pairing mechanism, may
be found in section \ref{kamsec:string-theory-pict}.

The DBI action determines the shape of the brane embeddings, {i.\,e.\,} the scalar
fields $\phi$, as well as the configuration of the gauge fields $A$ on these
branes. We consider the case of $N_f=2$ coincident D$7$-branes for which the
non-Abelian DBI action reads \cite{kamMyers:1999ps}
\begin{equation}
 \label{kameq:non-AbelianDBI}
 S_{\mathrm{DBI}}=-T_{D7}\:\mathrm{Str}\int\!\mathrm{d}^8\xi\:\sqrt{\det
     Q}\Bigg[\det\Big(P_{ab}\big[E_{\mu\nu}+E_{\mu i}(Q^{-1}-\delta)^{ij}E_{j\nu}\big]+2\pi\alpha'F_{ab}\Big)\Bigg]^{\frac{1}{2}}
\end{equation}
with
\begin{equation}
 \label{kameq:defQDBI}
 Q^i{}_j=\delta^i{}_j+i 2\pi\alpha'[\Phi^i,\Phi^k]E_{kj}
\end{equation}
and $P_{ab}$ the pullback to the D$p$-brane, where for a D$p$-brane in $d$ dimensions we have 
$\mu,\,\nu=0,\dots, (d-1)$, $a,\,b=0,\dots, p$, $i,\,j = (p+1),\dots, (d-1)$,
$E_{\mu\nu} = g_{\mu\nu} + B_{\mu\nu}$. In our case we set $p=7$, $d=10$,
$B\equiv 0$. As in \cite{kamErdmenger:2008yj} we can simplify this action
significantly by using the spatial and gauge symmetries present in our setup.
The action becomes
\begin{eqnarray}
  \label{kameq:DBI}
    S_{\mathrm{DBI}}&=&-T_{D7}\int\!\mathrm{d}^8\xi\:\mathrm{Str}\sqrt{|\det(G+2\pi\alpha'F)|}\\
    &=&-T_{D7}\int\!\mathrm{d}^8\xi\:\sqrt{-G}\:\mathrm{Str}\Bigg[1+G^{00}G^{44}\left(F^3_{\varrho
        0}\right)^2\left(\sigma^3\right)^2+G^{33}G^{44}
    \left(F^1_{\varrho 3}\right)^2\left(\sigma^1\right)^2\nonumber\\
     &&+G^{00}G^{33}\left(F^2_{03}\right)^2\left(\sigma^2\right)^2\Bigg]^{\frac{1}{2}}\,,
\end{eqnarray}
where in the second line the determinant is calculated. Due to the
symmetric trace, all commutators between the matrices $\sigma^i$ vanish. It is
known that the symmetrized trace prescription in the DBI action is only valid
up to fourth order in $\alpha'$ \cite{kamTseytlin:1997csa,kamHashimoto:1997gm}. However the corrections to the higher order
terms are suppressed by $N_f^{-1}$ \cite{kamConstable:1999ac} (see also \cite{kamMyers:2008me}). Here we use two different approaches to
evaluate the non-Abelian DBI action (\ref{kameq:DBI}). First, we modify the
symmetrized trace prescription by omitting the commutators of the generators
$\sigma^i$ and then setting $(\sigma^i)^2=1$ (see subsection
\ref{kamsec:change-str-prescr} below). This prescription makes the calculation of
the full DBI action feasible. This prescription is not verified in general but we obtain
physically reasonable results as discussed in section~\ref{kamsec:thermoBroken} and \ref{kamsec:fluctuationsBroken}. 
Second, we expand the non-Abelian DBI
action to fourth order in the field strength $F$ (see subsection \ref{kamsec:expansion-dbi-action}).
Here it should be noted that in general the higher terms of this expansion
need not be smaller than the leading ones. However, we again get physical results in our specific case which
confirm this approach. We further motivate the validity of our two approaches below.
%
\subsubsection{Adapted symmetrized trace prescription}
\label{kamsec:change-str-prescr}
Using the adapted symmetrized trace prescription defined above, the action becomes
 \begin{eqnarray}
   \label{kameq:DBIchangedstr}
     S_{\mathrm{DBI}}&=&-T_{D7}N_f\int\!\mathrm{d}^8\xi\:\sqrt{-G}\Bigg[1+G^{00}G^{44}\left(F^3_{\varrho 0}\right)^2+G^{33}G^{44}\left(F^1_{\varrho 3}\right)^2+G^{00}G^{33}
     \left(F^2_{03}\right)^2\Bigg]^{\frac{1}{2}}\nonumber\\
     &=&-\frac{T_{D7}N_f}{4}\!\int\!\mathrm{d}^8\xi\;\varrho^3f\tilde{f}(1-\chi^2)\Upsilon(\rho,\chi,\tilde{A})\,,
 \end{eqnarray}
with
\begin{eqnarray}
  \label{kameq:Upsilon}
    \Upsilon(\rho,\chi,\tilde{A})=\Bigg[&1-\chi^2+\rho^2(\partial_\rho\chi)^2-\frac{2\tilde{f}}{f^2}(1-\chi^2)\left(\partial_\rho
      \tilde{A}^3_0\right)^2+\frac{2}{\tilde{f}}(1-\chi^2)\left(\partial_\rho \tilde{A}^1_3\right)^2\nonumber\\
    &-\frac{2\gamma^2}{\pi^2\rho^4f^2}(1-\chi^2+\rho^2(\partial_\rho\chi)^2)
    \left(\tilde{A}^3_0\tilde{A}^1_3\right)^2\Bigg]^{\frac{1}{2}}\,,
\end{eqnarray}
where the dimensionless quantities $\rho=\varrho/\varrho_H$ and
$\tilde{A}=(2\pi\alpha')A/\varrho_H$ are used. To obtain first order equations of
motion for the gauge fields which are easier to solve numerically, we perform a
Legendre transformation. Similarly to
\cite{kamKobayashi:2006sb,kamErdmenger:2008yj} we calculate the electric displacement
$p_0^3$ and the magnetizing field $p_3^1$ which are given by the conjugate
momenta of the gauge fields $A_0^3$ and $A_3^1$,
\begin{equation}
  \label{kameq:conmomenta}
  p_0^3=\frac{\delta S_{\mathrm{DBI}}}{\delta(\partial_\varrho A^3_0)}\,,\qquad
  p_3^1=\frac{\delta S_{\mathrm{DBI}}}{\delta(\partial_\varrho A_3^1)}\,.
\end{equation}
In contrast to
\cite{kamKobayashi:2006sb,kamKarch:2007br,kamMateos:2007vc,kamErdmenger:2008yj}, the
conjugate momenta are not constant any more but depend on the radial coordinate
$\varrho$ due to the non-Abelian term $A_0^3A_3^1$ in the DBI action. For the
dimensionless momenta $\tilde{p}_0^3$ and $\tilde{p}_3^1$ defined as  
\begin{equation}
  \label{kameq:pt}
  \tilde{p}=\frac{p}{2\pi\alpha'N_fT_{D7}\varrho_H^3}\,,
\end{equation}
we get
\begin{equation}
  \label{kameq:conmomentadim}
  \tilde{p}_0^3=\frac{\rho^3\tilde{f}^2(1-\chi^2)^2\partial_\rho\tilde{A}_0^3}{2f\Upsilon(\rho,\chi,\tilde{A})}\,,\quad\tilde{p}_3^1=-\frac{\rho^3f(1-\chi^2)^2\partial_\rho\tilde{A}_3^1}{2\Upsilon(\rho,\chi,\tilde{A}))}\,.
\end{equation}
Finally, the Legendre-transformed action is given by
\begin{eqnarray}
  \label{kameq:DBIlegendre}
      \tilde S_{\mathrm{DBI}}&=S_{\mathrm{DBI}}-\int\!\mathrm{d}^8\xi\:
    \Bigg[
      \left(\partial_\varrho A_0^3\right)\frac{\delta S_{\mathrm{DBI}}}{\delta
        \left(\partial_\varrho A_0^3\right)}+\left(\partial_\varrho A_3^1\right)\frac{\delta S_{\mathrm{DBI}}}{\delta\left(\partial_\varrho A_3^1\right)}
    \Bigg]\\
    &=-\frac{T_{D7}N_f}{4}\int\!\mathrm{d}^8\xi\:\varrho^3f\tilde{f}(1-\chi^2)\sqrt{1-\chi^2+\rho^2(\partial_\rho\chi)^2}\;V(\rho,\chi,\tilde{A},\tilde{p})\,,
\end{eqnarray}
with
\begin{eqnarray}
  \label{kameq:V}
  V(\rho,\chi,\tilde{A},\tilde{p})&=&\Bigg[
  \left(1-\frac{2\gamma^2}{\pi^2\rho^4f^2}\left(\tilde{A}_0^3\tilde{A}_3^1\right)^2\right)\nonumber\\
  &&\times\Bigg(1+\frac{8
    \left(\tilde{p}_0^3\right)^2}{\rho^6\tilde{f}^3(1-\chi^2)^3}
    -\frac{8
    \left(\tilde{p}_3^1\right)^2}{\rho^6\tilde{f} f^2(1-\chi^2)^3}\Bigg)\Bigg]^{\frac{1}{2}}\,.
\end{eqnarray}
Then the first order equations of motion for the gauge fields and their conjugate momenta are
\begin{eqnarray}
  \label{kameq:eomgauge}
    \partial_\rho\tilde{A}_0^3&=\frac{2f\sqrt{1-\chi^2+\rho^2(\partial_\rho\chi)^2}}{\rho^3\tilde{f}^2(1-\chi^2)^2}\tilde{p}_0^3W(\rho,\chi,\tilde{A},\tilde{p})\, , \\
    \partial_\rho\tilde{A}_3^1&=-\frac{2\sqrt{1-\chi^2+\rho^2(\partial_\rho\chi)^2}}{\rho^3f(1-\chi^2)^2}\tilde{p}_3^1W(\rho,\chi,\tilde{A},\tilde{p})\, , \\
    \partial_\rho\tilde{p}_0^3&=\frac{\tilde{f}(1-\chi^2)\sqrt{1-\chi^2+\rho^2(\partial_\rho\chi)^2}c^2}{2\pi^2\rho
      fW(\rho,\chi,\tilde{A},\tilde{p})}\left(\tilde{A}_3^1\right)^2\tilde{A}_0^3\, , \\
    \partial_\rho\tilde{p}_3^1&=\frac{\tilde{f}(1-\chi^2)\sqrt{1-\chi^2+\rho^2(\partial_\rho\chi)^2}c^2}{2\pi^2\rho
      fW(\rho,\chi,\tilde{A},\tilde{p})}\left(\tilde{A}_0^3\right)^2\tilde{A}_3^1\,,
\end{eqnarray}
with
\begin{equation}
  \label{kameq:W}
   W(\rho,\chi,\tilde{A},\tilde{p})=\sqrt{\frac{1-\frac{2\gamma^2}{\pi^2\rho^4f^2}\left(\tilde{A}_0^3\tilde{A}_3^1\right)^2}{1+\frac{8
      \left(\tilde{p}_0^3\right)^2}{\rho^6\tilde{f}^3(1-\chi^2)^3}-\frac{8
      \left(\tilde{p}_3^1\right)^2}{\rho^6\tilde{f}
      f^2(1-\chi^2)^3}}}\,.
\end{equation}
For the embedding function $\chi$ we get the second order equation of motion
\begin{eqnarray}
  \label{kameq:eomchi}
    \partial_\rho
    \left[\frac{\rho^5f\tilde{f}(1-\chi^2)(\partial_\rho\chi)}{\sqrt{1-\chi^2+\rho^2(\partial_\rho\chi)^2}}V\right]
    =&-\frac{\rho^3f\tilde{f}\chi}{\sqrt{1-\chi^2+\rho^2(\partial_\rho\chi)^2}}\Bigg[
    \left[3\left(1-\chi^2\right)+2\rho^2(\partial_\rho\chi)^2\right]V\nonumber\\
    &-\frac{24\left(1-\chi^2+\rho^2(\partial_\rho\chi)^2\right)}{\tilde{f}^3\rho^6\left(1-\chi^2\right)^3}W
    \left(\left(\tilde{p}_0^3\right)^2-\frac{\tilde{f}^2}{f^2}\left(\tilde{p}_3^1\right)\right)\Bigg]\,.
\end{eqnarray}
We solve the equations numerically and determine the solution by integrating 
the equations of motion
from the horizon at $\rho=1$ to the boundary at $\rho=\infty$. The initial
conditions may be determined by the asymptotic expansion of the gravity fields near the
horizon
\begin{eqnarray}
  \label{kameq:asymh}
    \tilde{A}_0^3&= &\frac{c_0}{\sqrt{(1-\chi_0^2)^3+c_0^2}}(\rho-1)^2+{\cal O}\left((\rho-1)^3\right) \, , \\
    \tilde{A}_3^1&=b_0& +{\cal O}\left((\rho-1)^3\right) \,  , \\
    \tilde{p}_0^3&=c_0&+\frac{\gamma^2b_0^2c_0}{8\pi^2}(\rho-1)^2+{\cal O}\left((\rho-1)^3\right) \, , \\
    \tilde{p}_3^1&=& +{\cal O}\left((\rho-1)^3\right) \, , \\
    \chi&=\chi_0&-\frac{3\chi_0(1-\chi_0^2)^3}{4[(1-\chi_0^2)^3+c_0^2]}(\rho-1)^2+{\cal O}\left((\rho-1)^3\right)\, ,
\end{eqnarray}
where
the terms in the expansions 
are arranged according to their order in $\rho-1$.
For the numerical calculation we consider the terms up to sixth order in $\rho - 1$. The
three independent parameter $b_0$, $c_0$ and $\chi_0$ may be determined by
field theory quantities defined via the asymptotic expansion of the gravity
fields near the boundary,
\begin{eqnarray}
  \label{kameq:asymb}
    \tilde{A}_0^3&=\tilde{\mu}& -\frac{\tilde{d}_0^3}{\rho^2}+{\cal O}\left(\rho^{-4}\right) \, , \\
    \tilde{A}_3^1&=& -\frac{\tilde{d}_3^1}{\rho^2} +{\cal O}\left(\rho^{-4}\right)\, , \\
    \tilde{p}_0^3&=\tilde{d}_0^3& +{\cal O}\left(\rho^{-4}\right) \, , \\
    \tilde{p}_3^1&=-\tilde{d}_3^1& +\frac{\gamma^2 \tilde{\mu}^2\tilde{d}_3^1}{4 \pi^2\rho^2} +{\cal O}\left(\rho^{-4}\right) \, , \\
    \chi&= &\frac{m}{\rho}+\frac{c}{\rho^3}+{\cal O}\left(\rho^{-4}\right)\,.
\end{eqnarray}
According to the AdS/CFT dictionary, $\mu$ is the isospin chemical potential.
The parameters $\tilde{d}$ are related to the vev of the flavor currents $J$ by
\begin{equation}
  \label{kameq:dt}
  \tilde{d}_0^3=\frac{2^{\frac{5}{2}}\langle
    J_0^3\rangle}{N_fN_c\sqrt{\lambda}T^3}\,,\quad \tilde{d}_3^1=\frac{2^{\frac{5}{2}}\langle J_3^1\rangle}{N_fN_c\sqrt{\lambda}T^3}
\end{equation}
and $m$ and $c$ to the bare quark mass $M_q$ and the quark condensate
$\langle\bar\psi\psi\rangle$, 
\begin{equation}
  \label{kameq:cm}
  m=\frac{2M_q}{\sqrt{\lambda}T}\,,\quad c=-\frac{8\langle\bar\psi\psi\rangle}{\sqrt{\lambda}N_fN_cT^3}\, , 
\end{equation}
respectively. There are two independent physical parameters, {e.\,g.\,} $m$ and $\mu$, in the grand
canonical ensemble. From the boundary asymptotics (\ref{kameq:asymb}), we also
obtain that there is no source term for the current $J_3^1$. Therefore
as a non-trivial result we find that the
$U(1)_3$ symmetry is always broken spontaneously. In contrast, 
in the related works on
p-wave superconductors in $2+1$ dimensions
\cite{kamGubser:2008wv,kamRoberts:2008ns}, the spontaneous breaking of the $U(1)_3$
symmetry has to be put in by hand by setting the source term for the
corresponding operator to zero. With the constraint $\tilde{A}_3^1|_{\rho\to\infty}=0$ and the two
independent physical parameters, we may fix the three independent parameters of
the near-horizon asymptotics and obtain a solution to the equations of motion.

\subsubsection{Expansion of the DBI action}
\label{kamsec:expansion-dbi-action}
We now outline the second approach which we use. Expanding the action
(\ref{kameq:DBI}) to fourth order in the field strength $F$ yields
 \begin{equation}
   \label{kameq:DBIaxpand}
   S_{\mathrm{DBI}}=-T_{D7}N_f\int\!\mathrm{d}^8\xi\:\sqrt{-G}
   \left[1+\frac{{\cal T}_2}{2}-\frac{{\cal T}_4}{8}+\cdots\right]\,,
 \end{equation}
where ${\cal T}_i$ consists of the terms with order $i$ in $F$. To
calculate the ${\cal T}_i$, we use the following results for the symmetrized traces
\begin{eqnarray}
  \label{kameq:str}
    \,&2\sigma:\quad &\mathrm{Str}\left[\left(\sigma^i\right)^2\right]=N_f \, , \\
    \,&4\sigma: &\mathrm{Str}\left[\left(\sigma^i\right)^4\right]=N_f\,,\quad\mathrm{Str}\left[\left(\sigma^i\right)^2\left(\sigma^j\right)^2\right]=\frac{N_f}{3}\,,
\end{eqnarray}
where the indices $i,j$ are distinct. Notice that the symmetric trace of
terms with unpaired $\sigma$ matrices vanish, {e.\,g.\,}
$\mathrm{Str}[\sigma^i\sigma^j]=N_f\delta^{ij}$. The ${\cal T}_i$ are given in the
appendix of~\cite{kamAmmon:2009fe}.

To perform the Legendre transformation of the above action, we determine the
conjugate momenta as in (\ref{kameq:conmomenta}). However, we cannot easily solve these
equations for the derivative of the gauge fields since we obtain two coupled equations
of third degree. Thus we directly calculate the equations of motion for the
gauge fields on the D$7$-branes. The equations are given in the appendix of~\cite{kamAmmon:2009fe}.

To solve these equations, we use the same strategy as in the adapted symmetrized trace 
prescription discussed above. We integrate the equations of motion from the horizon at
$\rho=1$ to the boundary at $\rho=\infty$ numerically. The initial conditions
may be determined by the asymptotic behavior of the gravity fields near the
horizon
\begin{eqnarray}
  \label{kameq:asymhexp}
    \,&\tilde{A}_0^3=& a_2(\rho-1)^2+{\cal O}\left((\rho-1)^3\right)\,,\\
    \,&\tilde{A}_3^1=&b_0+{\cal O}\left((\rho-1)^3\right)\,,\\
    \,&\chi=&\chi_0+\frac{3 (a_2^4+4a_2^2-8)\chi_0}{4 (3 a_2^4+4 a_2^2+8)}(\rho-1)^2+{\cal O}\left((\rho-1)^3\right)\,.
\end{eqnarray}
For the numerical calculation we use the asymptotic expansion up to sixth order.
As in the adapted symmetrized trace prescription, there are again three independent parameters
$a_2,b_0,\chi_0$. Since we have not performed a Legendre transformation, we
trade the independent parameter $c_0$ in the asymptotics of the conjugate
momenta $\tilde{p}_0^3$ in the symmetrized trace prescription with the independent parameter
$a_2$ (cf. asymptotics in equation (\ref{kameq:asymh})). However, the
three independent parameters may again be determined in field theory quantities
which are defined by the asymptotics of the gravity fields near the boundary
\begin{eqnarray}
  \label{kameq:asymbexp}
    \,&\tilde{A}_0^3=&\mu -\frac{\tilde{d}_0^3}{\rho^2}+{\cal O}\left(\rho^4\right)\,,\\
    \,&\tilde{A}_3^1=& -\frac{\tilde{d}_3^1}{\rho^2}+{\cal O}\left(\rho^4\right)\,,\\
    \,&\chi=& \frac{m}{\rho}+\frac{c}{\rho^3}+{\cal O}\left(\rho^4\right)\,.
\end{eqnarray}
The independent parameters $\mu,\tilde{d}_0^3,\tilde{d}_3^1,m,c$ are given by field theory
quantities as presented in (\ref{kameq:dt}) and (\ref{kameq:cm}). Again we find that
there is no source term for the current $J_3^1$, which implies
spontaneous
symmetry breaking.  Therefore the independent
parameters in both prescriptions are the same and we can use the same strategy
to solve the equations of motion as described below (\ref{kameq:cm}). 

\emph{\runinhead{Exercise:}
In~\cite{kamBasu:2008bh} only the leading order of the action (\ref{kameq:DBIaxpand}) quadratic in field 
fluctuations was considered. For a specific chemical potential an analytic solution with non-zero
$A_0^3$ and $A_3^1$ can be found. Show that the analytic solution found there (adapted to our AdS${}_5$ case)
\begin{equation}
A^3_0=\mu (1-\rho^4)\, , \qquad A^1_3 = \epsilon \frac{\rho^4}{(1+\rho^4)^2} \, ,
\end{equation}
indeed solves the equations of motion derived from the action expanded to quadratic order for
the chemical potential $\mu=4$. $\epsilon$ is a constant formed from the coupling constant and 
the vacuum expectation value of the dual current $\langle J^1_3 \rangle$. 
Note: This exercise requires some work. See also~\cite{Herzog:2009ci} for an application of this solution. 
}

\section{D-brane thermodynamics \& spectrum}
\label{kamsec:braneThermoSpec}
In this section we briefly review the results obtained for the thermodynamics and the
spectrum of our setup with a baryonic, and later an isospin chemical potential.
\begin{figure}[b]
\sidecaption
 \includegraphics[width=0.6\textwidth]{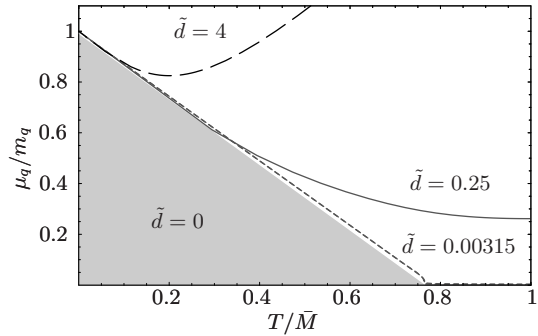}
 \caption{
   \label{kamfig:grandcanPhaseDiagB}
   The phase diagram in the canonical ensemble plotted against the variables of the grandcanonical ensemble. 
   On the axes the scaled chemical potential~$\mu_q/M_q$, with the quark mass~$M_q$
   is shown versus the scaled temperature~$T/\bar M= m^{-1}$ from~\cite{kamErdmenger:2007ja}.
 }
\end{figure}
%

\subsection{Baryon chemical potential}
\label{kamsec:baryonMu}
Figure \ref{kamfig:grandcanPhaseDiagB} shows the phase diagram of the field theory dual 
to the gravity setup we have constructed in the previous section \ref{kamsec:holoSetup}. In this
field theory we have introduced a baryonic chemical potential $\mu_q$ which is shown on the vertical axis
scaled by the quark mass $M_q$ defined by equation~(\ref{kameq:cm}). It is introduced by
a non trivial background gauge field solving the equation of motion derived from the DBI-action
and asymptoting to the chemical potential near the boundary as $\lim_{\rho\to\rho_B}A_t(\rho)=\mu_q$.
The horizontal axis shows the scaled temperature $T/\bar{M}=m^{-1}$. Recall that $m$
is the asymptotic value of the D7-brane embedding near the AdS boundary. In other words 
it is the source term for a quark condensate, or the non-normalizable mode of the brane embedding 
function. Here however we use it merely as a temperature scale. The lines in the diagram in 
figure~\ref{kamfig:grandcanPhaseDiagB} are lines of equal baryon density $\tilde d$. At low temperature
and chemical potential there is a triangle-shaped phase of zero density. Its diagonal borderline
to the white region with finite baryon density is the location of the so-called \index{meson melting transition}\emph{meson-melting transition}.
This is the transition where the fundamental matter melts, {i.\,e.\,} quark bound states, the
mesons, acquire an increasing decay width becoming quasi-particles. In the grey phase
we therefore have stable mesons with zero decay width. The deeper we go into the white phase
away from the transition line, the mesons melt. We will also see this in the spectral functions
computed below. As explained at the end of section~\ref{kamsec:flavor} we are going to stay 
entirely in the white phase at finite temperature where the brane embeddings are of black hole
type, which means that part of them falls into the black hole horizon (see section~\ref{kamsec:flavor}).

\subsubsection{Correlator recipe}
\label{kamsec:recipe}
We need to examine fluctuations of the fields in our setup in order to determine its 
spectrum and stability. There might be a fluctuation which is tachyonic and therefore 
could destabilize the whole system. For this purpose we quickly review how to compute
\index{correlator recipe}\index{real-time correlation functions}real-time correlation functions in gauge/gravity.

Let us work along the example of a gauge field fluctuation $a_\mu$. This appears in the
action, in our case the Dirac-Born-Infeld action, in the field strength $F=\D a + a\wedge a$.
Note that a background gauge field $A$ would appear in this field strength too, as we will see in later sections.
For simplicity we consider Abelian gauge field fluctuations without background $A$, 
{i.\,e.\,} $F_{\mu\nu}=\partial_\mu a_\nu-\partial_\nu a_\mu$. The action then reads something like
\begin{equation}
S\sim \int \D^8 \xi \sqrt{P[g_{\mu\nu}]+F_{\mu\nu}} \, ,
\end{equation}
with the $P$ being the pullback of the metric $g$ to the flavor brane. Expanding this action to
quadratic order in fluctuations $a$, we get the linearized equation of motion for example for
the spatial fluctuation $a_y$ which in Fourier space looks like this
\begin{equation}
\label{kameq:eoma}
0={\partial_\rho}^2 a_y + \frac{\partial_\rho [\sqrt{-g} g^{yy} g^{\rho\rho}]}{\sqrt{-g} g^{yy} g^{\rho\rho}} \partial_\rho a_y
 +\frac{g^{tt}}{g^{\rho\rho}} w^2 a_y\, ,
\end{equation}
where $w$ is the dimensionless frequency of the fluctuation. Also for other fields
we end up with a second order differential equation that we need to solve. Usually these equations
have singular (at the horizon $\rho_H$) coefficients which need to be regularized by an appropriate ansatz. In order to
find the most singular behavior solving this equation, we plug the ansatz $(\rho-\rho_H)^\beta$ into
equation~(\ref{kameq:eoma}) and expand. The leading order is a quadratic equation which can be solved for 
$\beta$ giving $\beta_{\mathrm{in}}$ or $\beta_{\mathrm{out}}$. Only one of those two solutions, $\beta_{\mathrm{in}}$ 
describes a fluctuation falling into the horizon, the other one is outgoing. We discard the outgoing one
because nothing is supposed to leave a classical black hole. $\beta$ is sometimes  called the 
\index{indicial exponent}\emph{indicial exponent}. Now we plug $a=(\rho-\rho_H)^{\beta_{\mathrm{in}}} F(\rho)$, with 
$F(\rho)=\sum\limits_{n=0}^\infty f_n (\rho-\rho_H)^n$
being a regular function of $\rho$, into the equation of motion (\ref{kameq:eoma}). Note that there
might be logarithmic terms present as explained in general for example in~\cite{kamBender} and 
discussed in detail in \cite{kamKaminski:2008ai}.
Picking the \index{infalling wave boundary condition}\emph{ingoing wave} fixes one of the two boundary conditions. The other boundary 
condition may be fixed by choosing the normalization of $F(\rho)$, {i.\,e.\,} the value $f_0$. 
All the higher $f_n$ depend recursively on $f_0$ and the indicial exponent $\beta$.
Now we solve the equation for $F$ analytically or numerically. The correlator is then 
obtained from the quadratic part of the on-shell action, which in our case has the structure
$S_{\mathrm{on-shell}}\sim \int \D \rho B(\rho) a \partial_\rho a$, where $B$ is a function 
of $\rho$ depending on metric coefficients. The recipe developed in~\cite{kamSon:2002sd,kamHerzog:2002pc}
tells us to strip the boundary values $a^{\mathrm{bdy}}$ off from the fields $a=a^{\mathrm{bdy}} {\cal A}$ and identify
what remains of the integrand with the Green function at the boundary $\rho=\rho_B$
\begin{equation}
G_{\mu\nu}^R= \lim\limits_{\rho\to\rho_B} B(\rho) {{\cal A}_\mu \partial_\rho {\cal A}_\nu}
 \, .
\end{equation} 
Now we only need to plug in our solutions $a=a^{\mathrm{bdy}} {\cal A}$. More details on the analytic and
numerical procedures are explained in \cite{kamKaminski:2008ai}.

\subsubsection{Spectral function \& quasi-normal modes}
\label{kamsec:specFuncB}
\begin{figure}[b]
  \sidecaption
  \includegraphics[width=0.6\linewidth]{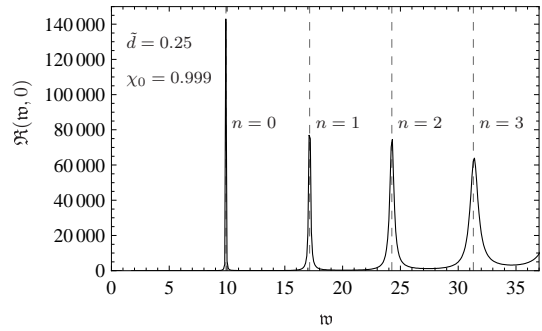}
  \caption{The spectral function ${\cal R}$ (in units of~$N_f N_c
			T^2/4$) at finite baryon density~$\tilde d$ versus dimensionless 
			frequency~$w$. At large $\chi_0$
			here the peaks approach the dashed drawn line spectrum given
			by~(\ref{kameq:susyVectorMesonM}).}
  \label{kamfig:lineSpectrumDt025heavy}
\end{figure}

The spectral function ${\cal R}$ is obtained from the imaginary part of the Green function 
${\cal R}= -2\mathrm{Im} G^R$. It encodes the spectrum of our thermal field theory. In particular
in figure \ref{kamfig:lineSpectrumDt025heavy} we are able to identify pretty stable 
quasi-particle excitations at a low finite temperature parametrize by $\chi_0$ and at finite 
baryon density $\tilde d=0.25$. Lowering the temperature these quasiparticles approach 
the line spectrum (\ref{kameq:susyVectorMesonM}) indicated here as dashed vertical lines.
The reason is that at lower temperature our theory restores its original supersymmetry. 
This also tells us that the vector quasi-particles we see in the thermal spectral function
are vector mesons. Our mesons melt in the finite temperature, finite density phase as mentioned above 
in the discussion of the phase diagram \ref{kamfig:grandcanPhaseDiagB}.

\runinhead{Quasi normal modes (QNM)}
The difference between the zero temperature line spectrum and the finite temperature
spectrum of quasi-particle excitations lies in the nature of the corresponding eigenmodes.
In the zero temperature case there is no black hole, thus no dissipation on the gravity side. 
The system has well-defined normal modes at real frequencies $w\in {\mathbf{R}}$. At finite temperature however
the (quasi) eigenfrequencies are complex $w\in{\mathbf{C}}$ owing to the dissipation
into the field theory plasma or in the dual gravity picture: dissipation into the black hole. The modes
traveling with these (quasi) eigenfrequencies are called \index{quasinormal mode~(QNM)}\emph{quasi-normal modes~(QNM)}.
We can roughly think of quasi-normal modes as being those solutions to the gravity fluctuation 
equations which vanish at the AdS boundary.
These QNMs have been found to be identical to the poles in the field theory Green function.
Therefore the location of the QNMs is closely related to the location of the peaks in the
spectral function. At least some of the quasi-normal modes create quasi-particle peaks in 
the spectral function. In the zero temperature limit we can think of those QNMs as approaching
the real frequency axis and reaching it in the limit, becoming real-valued. The corresponding
quasiparticles become stable which means line-shaped in the spectrum.
A more complete picture of QNMs is given in~\cite{kamBerti:2009kk}. Quasi normal
modes of our particular D3/D7 system are discussed in~\cite{Erdmenger:2009ce,Kaminski:2009dh}.
%
\subsection{Isospin chemical potential}
\label{kamsec:isospinMu}
\begin{figure}[b]
    \sidecaption
    \includegraphics[width=.63\linewidth]{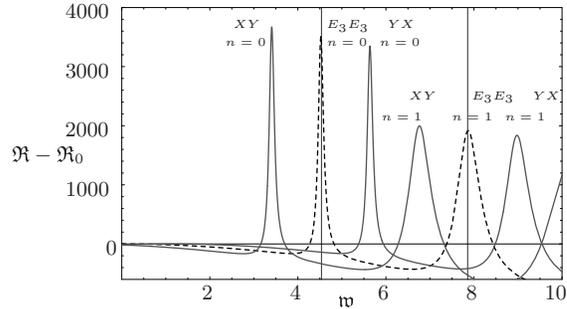}
    \caption{
        A comparison between the finite temperature part of the 
        spectral functions~${\cal R}_{XY}$ and ${\cal R}_{YX}$~(solid lines)
         in the two flavor directions~$X$ 
        and~$Y$ transversal to the chemical potential
        is shown in units of~$N_c T^2 T_r/4$ for large quark mass to temperature
        ratio~$\chi_0=0.99$ and~$\tilde d=0.25$. The spectral 
        function~${\cal R}_{E^3E^3}$ along the~$a=3$-flavor direction is shown as 
        a dashed line. 
    }
    \label{kamfig:isoXYsplitDt025}
\end{figure}
We can equally well introduce a chemical potential with $SU(2)$ --let's call it isospin-- structure,
represented by the Pauli matrices $\sigma^i$.
Choosing the chemical potential to point along the third flavor direction this background 
boils down to having two copies of the Abelian background gauge field $A_t(\rho)$ explored above
in the following way
\begin{equation}
\mu^3 =A_t(\rho) \sigma^3 = 
\left (
\begin{array}{cc}
A_t(\rho) & 0 \\
0 & -A_t(\rho)
\end{array}
\right ) \, .
\end{equation}
However we get some interesting new signatures through the $SU(2)$ structure. For example
the spectrum shown in figure~\ref{kamfig:isoXYsplitDt025} shows a triplet splitting for our mesons.
In particular we observe a splitting of the line expected at the lowest meson mass 
at~$w=4.5360$~($n=0$). The resonance is shifted to lower frequencies for~${\cal R}_{XY}$ 
and to higher ones for~${\cal R}_{YX}$, while it remains in place
for ~${\cal R}_{E^3E^3}$. The second meson resonance peak~($n=1$)
shows a similar behavior. So the different flavor combinations
propagate differently and have distinct quasi-particle resonances. This behavior is 
analogous to that of QCD's $\rho$ meson which is vector meson and a triplet under 
the isospin $SU(2)$ of QCD. Thus we have modeled the melting process of vector mesons
in a quark gluon plasma at finite isospin density.

\subsection{Instabilities \& the new phase}
\label{kamsec:instabilities}
What does all this have to do with our condensed matter motivation? The crucial thing is that
the isospin setup described above develops an instability at large enough isospin density. 
This means that the modes $X$ and $Y$ develop quasi-normal modes~(QNM) which 
have a positive imaginary part, {i.\,e.\,} which are enhanced instead of being damped.
This suggests that exactly the mesons corresponding to the $X,\, Y$ fluctuations condense.
This instability comes about naturally if we think about the fact, that we are trying to 
push more and more charge density into a confined volume. In particular the 3--7 strings 
charging the D7-brane are located at the black hole horizon as motivated earlier in section \ref{kamsec:flavor}.
Our current setup does not allow the strings to move into the bulk because we are forcing
all the background fields they would create there to be zero. Up to now we have required
$A^3_0\not =0$ and $A^a_\mu\equiv 0$ for all other field components. But we are going to
relax that restriction by allowing a non-trivial $A_3^1$. We will see below that this is sufficient to stabilize
the theory in a new phase which we will prove to be superconducting/superfluid. This setup
naturally produces a p-wave structure since we have seen above that the condensing 
vector mesons have a triplet structure. According to table \ref{kamtab:superCStates} this implies a
p-wave (at least to leading order).

\section{Signatures of super-something}
\label{kamsec:signatures}
Finally we put together everything we have learned about D-branes, superconductors and
holographic methods. We discuss the holographic results and interprete them in the 
condensed matter context. Section~\ref{kamsec:thermoBroken} starts with the thermodynamics 
and section~\ref{kamsec:fluctuationsBroken} continues with the details of the fluctuation
computation. The spectrum and conductivity is examined in section~\ref{kamsec:conductivityNSpec} and all 
the signatures will be pointing to the fact that \emph{we have indeed created a holographic
p-wave superconductor/superfluid}.

\subsection{Thermodynamics of the broken phase}
\label{kamsec:thermoBroken}
\begin{figure}[b]
  \centering
  \includegraphics[width=0.9\linewidth]{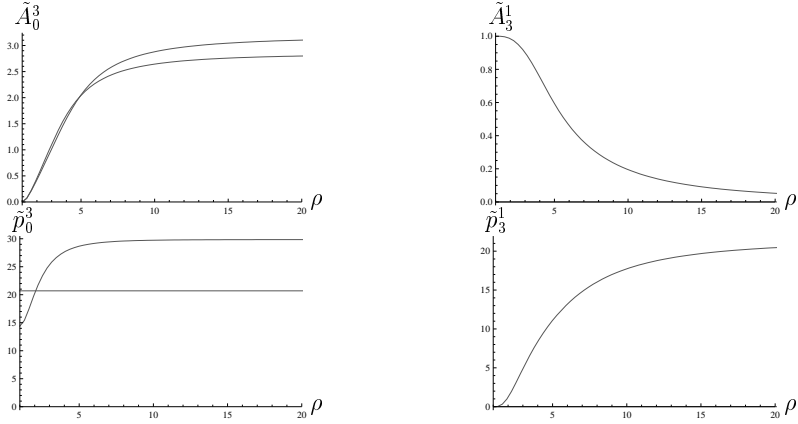}
  \caption{Profiles of the relevant dimensionless gauge fields $\tilde{A}$ on the D$7$-branes and
    their dimensionless conjugate momenta $\tilde{p}$ versus the dimensionless AdS radial coordinate
    $\rho$ near the horizon at $\rho=1$. (top left) the lower curve is $T=0.9 T_c$, 
    (bottom left) the constant is at $T=T_c$ since there are no charges in the bulk, the
    other curve shows that the charge proportional to $\tilde p^3_0$ increases
    towards the boundary.}
  \label{kamfig:profilesgaugefields}
\end{figure}
Figure \ref{kamfig:profilesgaugefields} shows the background field configuration.
The different curves correspond to the temperatures $T=T_c$ and $T\approx 0.9T_c$. The plots
are obtained at zero quark mass $m=0$ and by using the adapted
symmetrized trace prescription. Similar plots may also be obtained at
finite mass $m\not=0$ and by using the DBI action expanded to fourth
order in $F$. These plots show the same features: (top left) The gauge field $\tilde{A}_0^3$ increases monotonically
towards the boundary. At the boundary, its value is given by the
dimensionless chemical potential $\tilde{\mu}$. (top right) The gauge field $\tilde{A}_3^1$ is zero for $T\ge T_c$. For
$T<T_c$, its value is non-zero at the horizon and decreases monotonically
towards the boundary where its value has to be zero. (bottom left) The conjugate momentum
$\tilde{p}_0^3$ of the gauge field $\tilde{A}_0^3$ is constant for $T\ge T_c$. For
$T<T_c$, its value increases monotonically towards the boundary. Its boundary
value is given by the dimensionless density $\tilde{d}_0^3$. (bottom right) The conjugate
momentum $\tilde{p}_3^1$ of the gauge field $\tilde{A}_3^1$ is zero for $T\ge T_c$. For
$T<T_c$, its value increases monotonically towards the boundary. Its boundary
value is given by the dimensionless density $-\tilde{d}_3^1$.

All thermodynamic quantities are determined in terms of the relevant thermodynamic potential.
We use the grand potential $W_7$ in the grandcanonical and the free energy $F_7$ in the canonical ensemble.
Both stem from the D7-brane action. They are related through a Legendre transformation in the background gauge field $A^3_0$ on 
the gravity side. We obtain both potentials due to the gauge/gravity dictionary from the Euclideanized
gravity on-shell action according to $Z=\E^{-I_{\mathrm{on-shell}}}$, with the partition function $Z$ 
of the boundary field theory. So the for example $W_7= T S^{\mathrm{euclideanized}}_{\mathrm{D7, on-shell}}$.
\begin{figure}[b]
\centering   
  \includegraphics[width=0.85\linewidth]{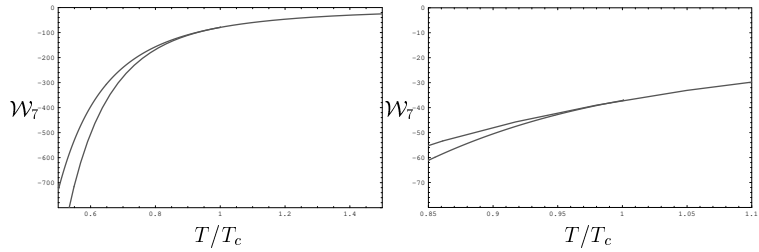} 
  \caption{Grand canonical potential computed {\bf (a)} from the symmetrized trace prescription
  and {\bf (b)} from the expanded DBI action, both
  at vanishing quark mass $M_q=0$. The qualitative behavior agrees
  for both prescriptions: At $T=T_c$ the energy curve splits
  into a lower and a higher energy branch. The branch with lower energy in both 
  cases is the one with a finite new condensate $\tilde{d}_3^1$ below $T_c$.
   } 
 \label{kamfig:W7dbiExp}
\end{figure}
Comparison of the grand potentials in figure~\ref{kamfig:W7dbiExp} 
then shows that the new phase with a finite value for $A_3^1$
is thermodynamically preferred below a temperature $T=T_c$ and does not exist above.
The transition seems numerically smooth, {i.\,e.\,} it is a \index{continuous phase transition}\emph{continuous phase transition}.
In these respects both computation schemes, the symmetrized trace prescription as well as the DBI expansion
to fourth order give the same qualitative behavior. 
\begin{figure}[b]
\sidecaption
  \includegraphics[width=0.6\linewidth]{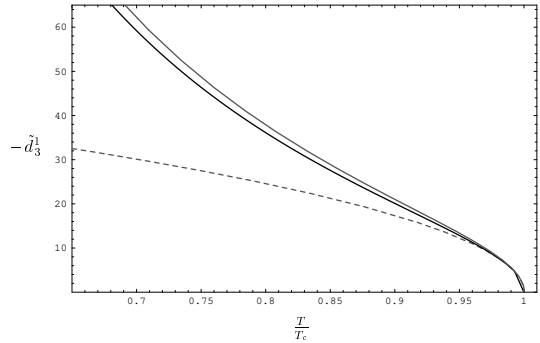}  
  \caption{The order parameter $\tilde{d}_3^1$ defined in (\ref{kameq:asymb}) as obtained from the 
  adapted symmetrized trace prescription versus temperature: The
  case of vanishing quark mass (top solid curve) shows the same behavior
  near $T_c$ as that where $\mu/M_q=3$ is fixed (lower solid curve). 
  Both curves go to zero with a critical exponent of $1/2$ near $T_c$,
  as visualized by the fit $55 (1-T/T_c)^{1/2}$ (dashed curve).
   } 
 \label{kamfig:d13symTr}
\end{figure}
Figure~\ref{kamfig:d13symTr} confirms this result and at least numerically determines the transition to be 
of second order. The order parameter $\tilde d^1_3$ vanishes with a critical exponent of $1/2$ to
numerical accuracy. It has been explicitly verified that this condensation of vector particles
occurs before, i.\,e.\, at higher temperature than the condensation of the scalars
in our theory (see remarks in~\cite{kamAmmon:2009fe}). If this was QCD the scalar pions would
condense before the vector mesons did. However, there might be other instabilities and also
other symmetry breaking configurations may be possible.
\begin{figure}[b]
\sidecaption
  \includegraphics[width=0.55\linewidth]{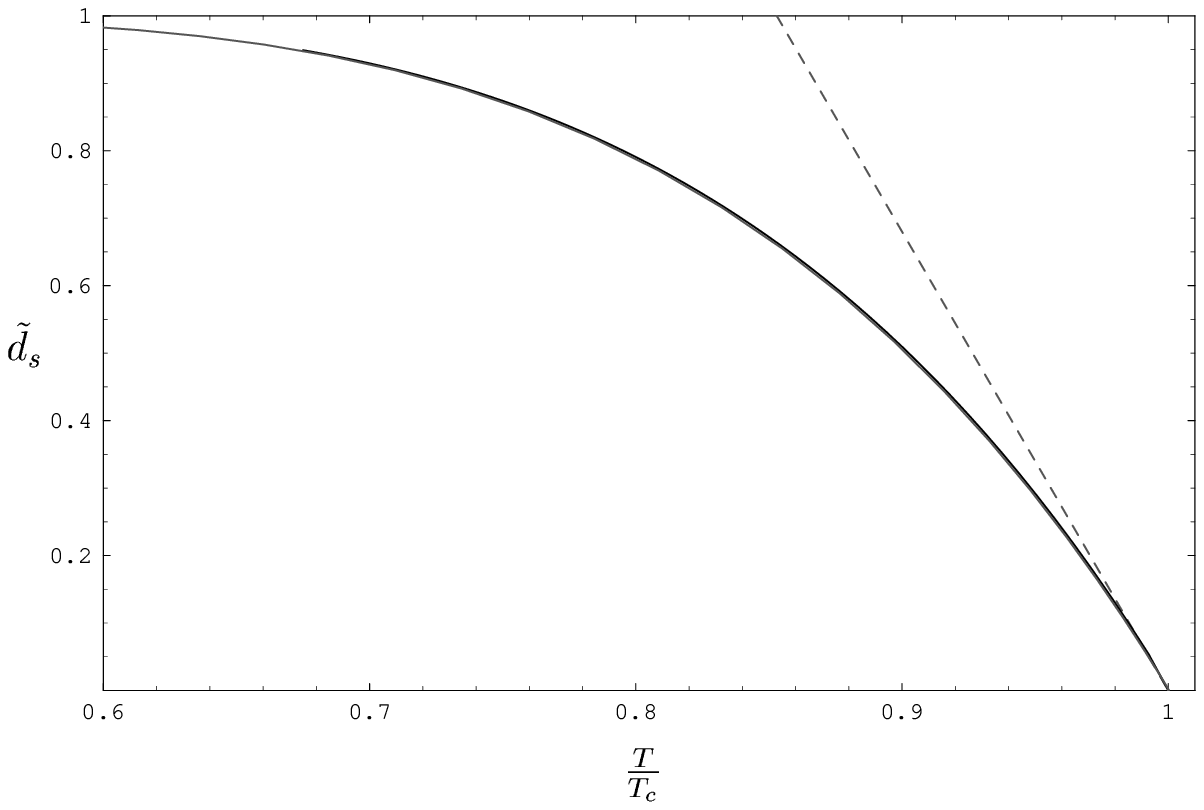}  
  \caption{Superconducting density $\tilde{d}_s=(\tilde{d}_0^3-c_0)/\tilde{d}_0^3$ versus temperature $T$:
  In both, the massless and the massive case at $\mu/M_q=3$ (both curves coincide on the plot), 
  the superconducting density $\tilde{d}_s$
  vanishes linearly at the critical temperature. This is visualized by the 
  fit $6.8 (1-T/T_c)$ (dashed).
   } 
 \label{kamfig:dssymTr}
\end{figure}

Using some intuition to define the superconducting density $\tilde{d}_s=(\tilde{d}_0^3-c_0)/\tilde{d}_0^3$, we
find that it vanishes linearly at the critical temperature as shown in figure \ref{kamfig:dssymTr}.
\begin{svgraybox}
All these signatures are those of a superconducting/superfluid phase transition. The
order parameter $\tilde d^1_3$ has vector structure by construction,  which implies the
p-wave.
\end{svgraybox}
\begin{figure}[b]
\sidecaption
  \includegraphics[width=0.55\linewidth]{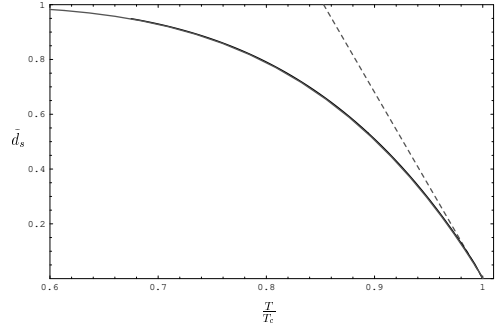}  
  \caption{The flavor brane contribution to the specific heat 
  as computed from the adapted symmetrized trace prescription in the massless
  case. The lower curve starting at $T=T_c$ is the heat capacity in the superconducting
  phase. The other curve is computed in the normal phase.
   } 
 \label{kamfig:CvsymTr}
\end{figure}
We can also compute the specific heat which the flavor branes contribute to the
theory as $C_7=-T {\partial}^2 F_7/{\partial T}^2$.
In figure~\ref{kamfig:CvsymTr} the higher curve corresponds to the normal phase with $A_3^1=0,$ while the
lower one corresponds to the superconducting phase with $A_3^1\not=0$.
Note that the total specific heat is always positive although the
flavor brane contribution is negative. The divergences near $T=0$ in both
phases can be attributed to the missing 
backreaction  in our setup\footnote{The backreaction
of the gauge field on the geometry, {i.\,e.\,} on the Einstein equations for metric
components has been considered in~\cite{kamAmmon:2009xh}.}. We read off
from the  numerical result that near the critical temperature, the dimensionless 
specific heat ${\cal C_7}=32 C_7/(\lambda N_f N_c T^3)$ is constant in the superconducting phase. 
\begin{svgraybox}
This implies that the dimensionful specific heat $C_7$ is proportional to $T^3$. This temperature
dependence is characteristic for Bose liquids. There is also a finite jump in the specific heat at $T_c$.
\end{svgraybox}

\subsection{Fluctuations in the broken phase}
\label{kamsec:fluctuationsBroken}
Let us now investigate fluctuations of this system. We are going to see that the computation
of the non-Abelian DBI-action is a very subtle issue. We argue for a novel prescription which
actually gives reasonable results. The full gauge field~$\hat A$ on the branes consists of the field~$A$ 
and fluctuations~$a$, 
\begin{equation} 
\hat A = A_0^3 \tau^3 \D t +A_3^1 \tau^1 \D x_3 +
  a_\mu^a \tau^a \D x^\mu \,,
\end{equation}
where $\tau^a$ are the $SU(2)$ generators. The linearized equations of motion
for the fluctuations $a$ are obtained by expanding the DBI action in $a$ to
second order. We will analyze the fluctuations $a^3_2$ and
$X=a^1_2+\I a^2_2$, $Y=a^1_2-\I a^2_2$.

Including these fluctuations, the DBI action reads 
\begin{equation}
\label{kameq:dbiFluctuations}  
S=-T_7\int \D^8\xi\: \mathrm{Str}\sqrt{\det{[G+(2\pi\alpha')\hat F]}} \,  , 
\end{equation}    
with the non-Abelian field strength tensor 
\begin{equation}
\hat F_{\mu\nu}^a = F_{\mu\nu}^a+\check F_{\mu\nu}^a \, ,  
\end{equation} 
where the background is collected in 
\begin{equation}  
F_{\mu\nu}^a = 2\partial_{[\mu} A_{\nu]}^a 
  +\frac{\gamma}{\sqrt{\lambda}} f^{abc} A_\mu^bA_\nu^c\, ,    
\end{equation}  
and all terms containing fluctuations in the gauge field are
summed in  
\begin{equation}  
\check F_{\mu\nu}^a = 2\partial_{[\mu} a_{\nu]}^a + 
   \frac{\gamma}{\sqrt{\lambda}} f^{abc} a_\mu^b a_\nu^c + 
   \frac{\gamma}{\sqrt{\lambda}} f^{abc} (A_\mu^b a_\nu^c+a_\mu^b A_\nu^c) \, .  
\end{equation}  
Index anti-symmetrization is always defined with a factor of two
in the following way $\partial_{[\mu}A_{\nu]}= (\partial_\mu A_\nu-\partial_\nu A_\mu)/2$.

\subsubsection{Adapted symmetrized trace prescription}
\label{kamsec:adopt-symm-trace}
In this section we use the adapted symmetrized trace prescription to determine
the fluctuations about the background we discussed in section
\ref{kamsec:change-str-prescr}. To obtain the linearized equations of motion for
the fluctuations $a$, we expand the action (\ref{kameq:dbiFluctuations}) to
second order in fluctuations, 
\begin{eqnarray}
    S^{(2)}&=&-T_7\int \D^8\xi\: \mathrm{Str}\Bigg[\sqrt{-{{\cal G}}}
      +\frac{(2\pi\alpha')}{2}\sqrt{-{{\cal G}}}{{\cal G}}^{\mu\nu}{{\check F }}_{\nu\mu}\\ \nonumber
      &&-\frac{(2\pi\alpha')^2}{4}
      \sqrt{-{{\cal G}}}
      {{\cal G}}^{\mu\mu'}{{\check F }}_{\mu'\nu}{{\cal G}}^{\nu\nu'}{{\check F }}_{\nu'\mu}
      +\frac{(2\pi\alpha')^2}{8}
      \sqrt{-{{\cal G}}}\left({{\cal G}}^{\mu\nu}{{\check F }}_{\nu\mu}\right)^2\Bigg] \,. 
\end{eqnarray}
As in \cite{kamErdmenger:2007ja}, we collect the metric and
gauge field background in the tensor ${{\cal G}}=G+(2\pi\alpha') F$. Using the
Euler-Lagrange equation, we get the linearized equation of motion for fluctuations~$a_\kappa^d$ in the form
\begin{eqnarray}
  \label{kameq:eomfluc}
    0&=\partial_\lambda\mathrm{Str}
    \Big[&\sqrt{-{{\cal G}}}\tau^d\Big\{{{\cal G}}^{[\kappa\lambda]}
    +(2\pi\alpha')\Big({{\cal G}}^{\mu[\kappa}{{\cal G}}^{\lambda]\nu}+\frac{1}{2}{{\cal G}}^{\mu\nu}{{\cal G}}^{[\kappa\lambda]}\Big)\check
        F_{\nu\mu}\Big\}\Big]\\ \nonumber
        &-\mathrm{Str}\Big[&\frac{c}{\sqrt{\lambda}}f^{abd}\tau^a\sqrt{-{{\cal G}}}
        \times\Big\{{{\cal G}}^{[\kappa\lambda]}(a+A)^b_\lambda+(2\pi\alpha')
          \Big({{\cal G}}^{\mu[\kappa}{{\cal G}}^{\lambda]\nu}\\ \nonumber
          &&+\frac{1}{2}{{\cal G}}^{\mu\nu}{{\cal G}}^{[\kappa\lambda]}\Big)\check F_{\nu\mu}A^b_\lambda\Big\}\Big]\,.
\end{eqnarray}
Note that the linearized version of the fluctuation field strength used in equation (\ref{kameq:eomfluc}) is given by 
\begin{equation}
  \check F_{\mu\nu}^a = 2\partial_{[\mu} a_{\nu]}^a +
  \frac{\gamma}{\sqrt{\lambda}} f^{abc} (A_\mu^b a_\nu^c+a_\mu^b A_\nu^c)
  +{\cal O} (a^2) \, .
\end{equation}
In our specific case the background tensor in its covariant form is given by
\begin{equation} 
{{\cal G}}_{\mu\nu}=G_{\mu\nu} \tau^0 + (2\pi\alpha')\Big ( 
 2\partial_\varrho A_0^3\delta_{4[\mu}\delta_{\nu]0}\tau^3+ 2\partial_\varrho A_3^1\delta_{4[\mu}\delta_{\nu]3}\tau^1
 + 2\frac{\gamma}{\sqrt{\lambda}}A_0^3A_3^1\delta_{0[\mu}\delta_{\nu]3}\tau^2
\Big ) \,. 
\end{equation}   
Inversion yields the contravariant form needed to compute the 
explicit equations of motion. The inverse of ${{\cal G}}$ is defined as
${{\cal G}}^{\mu\nu}{{\cal G}}_{\nu\mu'}=\delta^\mu_{\mu'}\tau^0$ \footnote{We calculate the
  inverse of ${{\cal G}}$ by ignoring the commutation relation of the $\tau$'s because of the symmetrized trace.
It is important that $\tau^a\tau^b$ must not be simplified to
$\epsilon^{abc}\tau^c$ since the symmetrization is not the same for these two expressions.}. The non-zero
components of ${{\cal G}}^{\mu\nu}$ may be found in the appendix of~\cite{kamAmmon:2009fe}.

\runinhead{Fluctuations in $a^3_2$:}
For the fluctuation $a_2^3$ with zero spatial momentum, we obtain the equation
of motion
\begin{eqnarray}
  \label{kameq:eoma32}
  0&=(a^3_2)''+\frac{\partial_\rho H}{H}(a^3_2)'-
  \Bigg[&\frac{4\varrho_H^4}{R^4}\left(\frac{{{\cal G}}^{33}}{{{\cal G}}^{44}}(\mathcal{M}_3^1)^2+\frac{{{\cal G}}^{00}}{{{\cal G}}^{44}}w^2\right)
  \nonumber \\
  &&-16\frac{\partial_\rho\left(\frac{H}{\rho^4f^2}\tilde{A}^3_0(\partial_\rho\tilde{A}^3_0)(\mathcal{M}^1_3)^2\right)}{H\left(1-\frac{2c^2}{\pi^2\rho^4f^2}(\tilde{A}^1_3\tilde{A}^3_0)^2\right)}\Bigg]a^3_2\,,
\end{eqnarray}
with $\mathcal{M}^1_3={\gamma\tilde{A}^1_3}/{(2\sqrt{2}\pi)}$ and $H=\sqrt{{{\cal G}}}G^{22}{{\cal G}}^{44}$.

\runinhead{Fluctuations in $X=a^1_2+\I a^2_2$, $Y=a^1_2-\I a^2_2$:}
For the fluctuations $X$ and $Y$ with zero spatial momentum, we obtain the
coupled equations of motion
\begin{eqnarray}
  \label{kameq:eomXY}
    0=&X''+\frac{\partial_\rho H}{H}X'-\frac{4\varrho_H^4}{R^4}
    \Bigg[\frac{{{\cal G}}^{00}}{{{\cal G}}^{44}}\left(w-\mathcal{M}^3_0\right)^2+\frac{{{\cal G}}^{\{03\}}}{{{\cal G}}^{44}}\mathcal{M}^1_3w\Bigg]X+\frac{4\varrho_H^4}{R^4}
    \Bigg[\frac{{{\cal G}}^{\{03\}}}{{{\cal G}}^{44}}\mathcal{M}^1_3\mathcal{M}^3_0\nonumber \\ \nonumber
    &+\frac{R^2}{4\varrho_H^2}\frac{\partial_\rho\left[\sqrt{-{{\cal G}}}G^{22}{{\cal G}}^{\{34\}}\mathcal{M}^1_3\right]}{H}-\frac{{{\cal G}}^{33}}{2{{\cal G}}^{44}}\left(\mathcal{M}^1_3\right)^2\Bigg](X-Y)+\frac{4\varrho_H^2}{R^2}\frac{{{\cal G}}^{\{04\}}}{{{\cal G}}^{44}}w 
    Y'\\
    &+\frac{2\varrho^2_H}{R^2}\frac{\partial_\rho
      \left[\sqrt{-{{\cal G}}}G^{22}{{\cal G}}^{\{04\}}\left(w+\mathcal{M}^3_0\right)\right]}{H}Y \, ,
\end{eqnarray}
where the component of the inverse background tensor may be found in the
appendix of~\cite{kamAmmon:2009fe} (just like the corresponding formula for $Y$ which is only different
from (\ref{kameq:eomXY}) by a few signs), index symmetrization is defined
${{\cal G}}^{\{ij\}}=({{\cal G}}^{ij}+{{\cal G}}^{ji})/2$ and $\mathcal{M}^3_0={\gamma\tilde{A}^3_0}/{(2\sqrt{2}\pi)}$.

\subsubsection{Expansion of the DBI action}
\label{kamsec:expansion-dbi-action-1}
In this section we determine the equation of motion for the fluctuation $a_2^3$
in the background determined by the DBI action expanded to fourth order in
$F$ (see section \ref{kamsec:expansion-dbi-action}). To obtain the
quadratic action in the field $a_2^3$, we first have to expand the DBI action
(\ref{kameq:dbiFluctuations}) to fourth order
in the full gauge field strength $\hat F$, and expand the result
to second order in $a_2^3$. Due to the symmetries of our setup, the equation
of motion for the fluctuation $a_2^3$ at zero spatial momentum decouples from
the other equations of motion, such that we can write down an effective
Lagrangian for the fluctuation $a_3^2$. This effective Lagrangian is given in
the appendix of~\cite{kamAmmon:2009fe}. The equation of motion for $a_2^3$
with zero spatial momentum determined by the Euler-Lagrange equation is
given by 
\begin{eqnarray}
  \label{kameq:eoma32exp}
  0=&(a_2^3)''+\frac{\partial_\rho {\cal H}}{{\cal H}}(a_2^3)'-
  \frac{\varrho_H^4}{R^4}\Bigg[4
    \Bigg(\frac{{\cal H}^{00}}{{\cal H}^{44}}w^2+\frac{{\cal H}^{33}}{{\cal H}^{44}}(\mathcal{M}_3^1)^2\Bigg)\\
    &+\frac{8}{3}\frac{\partial_\rho[\sqrt{-G}G^{00}G^{22}G^{33}G^{44}\tilde{A}_0^3(\partial_\rho
      \tilde{A}_0^3)(\mathcal{M}_3^1)^2]}{{\cal H}}\Bigg]a_2^3\,,
\end{eqnarray}
where ${\cal H}=\sqrt{-G}G^{22}{\cal H}^{44}$.
We introduce the factors ${\cal H}^{ij}$ which may be found in the appendix of~\cite{kamAmmon:2009fe}
to emphasize the similarity to the equation of
motion obtained by the adapted symmetrized trace prescription
(\ref{kameq:eoma32}).

\subsection{Conductivity \& spectrum}
\label{kamsec:conductivityNSpec}
\begin{figure}[b]
\centering
\includegraphics[width=.9\linewidth]{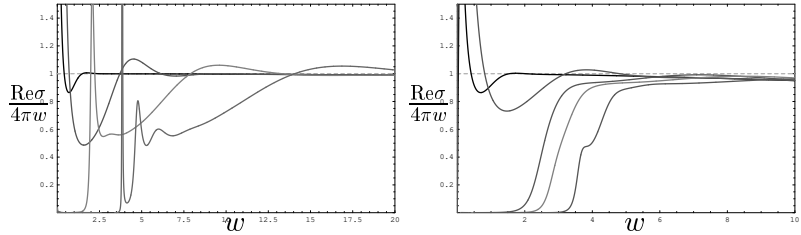}  
\caption{
Conductivity {\bf (a)} from the symmetrized trace prescription
at $T/T_c=10,\,1,\,0.5,\, 0.28$~(from left to right),
{\bf (b)} from the DBI action expanded to fourth order at
$T/T_c=\infty,\,1,\,0.6,\,0.5,\,0.39$.
}
\label{kamfig:conductivities}
\end{figure}
We calculate the frequency-dependent conductivity $\sigma(\omega)$ using the Kubo formula,
\begin{equation}
  \label{kameq:defsigma}
  \sigma(\omega)=\frac{\I}{\omega}G^R(\omega,q=0)\,,
\end{equation}
where $G^R$ is the retarded Green function of the current $J^3_2$ dual to
the fluctuation $a^3_2$, which we calculate using the method obtained
in~\cite{kamSon:2002sd}. The current $J^3_2$ is the analog to the electric
current since it is charged under the $U(1)_3$ symmetry. In real space it is
transverse to the condensate. Since this fluctuation is the only one which
transforms as a vector under the $SO(2)$ rotational symmetry, it decouples
from the other fluctuations of the system.
\emph{
\runinhead{Exercise:} Prove equation (\ref{kameq:defsigma}) for the current $J_2^3$ 
assuming that the gauge/gravity correspondence is correct. Recall that in regular 
electrodynamics $\sigma =J/E$ with the electric current $J$ and the electric field $E$.
Also recall that the two-point Green function for a current $J$ dual to the gauge field
$A$ can holographically be written in the form $G^R\propto \partial_\rho A / A$.
}
\begin{figure}[b]
\sidecaption
  \includegraphics[width=0.6\linewidth]{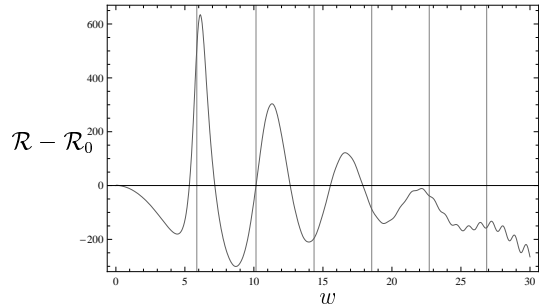}  
  \caption{Finite temperature part of the spectral function
    $\mathcal{R}-\mathcal{R}_0$ with $\mathcal{R}_0=4\pi w^2$in units of 
  $N_f N_c T^2/8$ versus the dimensionless frequency $w=\omega/(2\pi T)$ 
  at finite quark mass $m=2.842$ and chemical potential $\tilde{\mu}=3.483$. The grey lines
  correspond to the supersymmetric mass spectrum calculated in \cite{kamKruczenski:2003be}.
   } 
 \label{kamfig:RhighmasssymTr}
\end{figure}
\runinhead{(Pseudo) gap}
An analysis of the imaginary part of the conductivity using Kramers-Kronig realtions 
shows a delta peak at vanishing frequency $w=0$ in the real part. The frequency-dependent
conductivity is shown for our two distinct computation schemes in figure~\ref{kamfig:conductivities}.
Independent from the scheme we see an energy gap develop and grow while
the temperature is decreased. The temperature of the black hole horizon induced on the
D7-branes is proportional to the inverse particle mass parameter $m^{-1}\propto T$. Therefore
from the trivial flat brane embedding at $m=0$ we get an infinite temperature in 
subfigure \ref{kamfig:conductivities}~{\bf(b)}. Both schemes show the development of peaks
in the conductivities. The peaks coming from fluctuations around the symmetrized trace prescription
background are a lot more pronounced. Taking into account only second order terms in
the expanded DBI action as in~\cite{kamBasu:2008bh} would hide the peak structures completely.
Therefore we conclude that these peaks are higher order effects. Since the conductivity is
closely related to the spectral functions, we interprete the peaks as quasiparticles just as
described in section~\ref{kamsec:specFuncB} and~\ref{kamsec:isospinMu}. In particular 
these are again vector mesons. This identification is confirmed by the spectral function
in figure~\ref{kamfig:RhighmasssymTr}. The peaks in that figure are identical with those in
the conductivity and they approach the supersymmetric line spectrum for vector mesons, as
in section~\ref{kamsec:isospinMu}.

\runinhead{Dynamical mass generation}
Even if we choose the two D7 branes to coincide with the stack of D3-branes, i.\,e.\, if we
choose the quark mass to vanish, we observe the quasi-particle peaks mentioned before.
This is due to a Higgs-like mechanism which dynamically generates masses for the bulk
field fluctuations, which in turn give massive quasi-particles in the boundary theory. 
In the bulk our fields $A_0^3$ and $A_3^1$ break the $SU(2)$ symmetry
spontaneously since the bulk action is still $SU(2)$-invariant. Thus there are three
Nambu-Goldstone bosons which are immediately eaten by the bulk gauge fields, giving them
mass. This can be seen explicitly in the action where the following mass terms for the
gauge field fluctuations appear:$(A_0^3)^2 (a^{1,2})^2$ and $(A_3^1)^2 (a^3)^2$.
%
\subsection{Meissner-Ochsenfeld-Effect}
\label{kamsec:meissnerO}
The Meissner effect is a distinct signature of conventional 
and unconventional superconductors. It is the phenomenon of 
expulsion of external magnetic fields. 
An induced current in the superconductor generates a magnetic
field counter-acting the external magnetic field~$H$. In AdS/CFT 
we are not able to observe the generation of counter-fields since
the symmetries on the boundary are always global. Nevertheless,
we can study their cause, i.e. the current induced in the superconductor.
As usual
\cite{kamNakano:2008xc,kamAlbash:2008eh,kamMaeda:2008ir,kamHartnoll:2008kx}
the philosophy here is to weakly gauge the boundary theory afterwards.
\begin{figure}[b]
\sidecaption
  \includegraphics[width=0.6\linewidth]{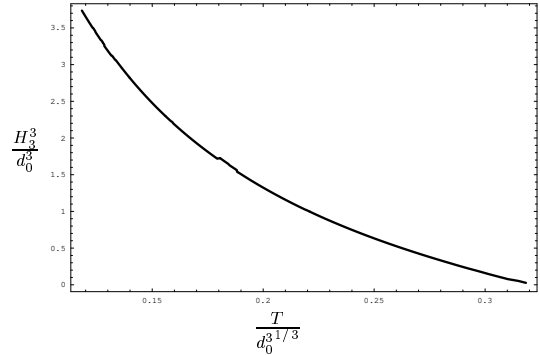}
  \caption{\label{kamfig:criticalBT}
    The line of critical magnetic field versus critical temperature. Below
    this line the external magnetic field coexists with the superconducting
    condensate. Above the line the superconducting condensate vanishes.
    We set the spatial position arbitrarily to $\tilde x=0.1$ since the critical
    line does not depend on $\tilde x$.
  }
\end{figure}
In order to investigate how an external magnetic field influences our
p-wave superconductor, we have two choices. Either we introduce the
field along the spatial $z$-direction $H_3^3\tau^3$ or equivalently one along the $y$-direction
i.e. $H_2^3\tau^3$. Both are ``aligned''~with the spontaneously broken
$U(1)_3$-flavor direction.

As an example here we choose a non-vanishing $H_3^3\tau^3$. This requires
inclusion of some more non-vanishing field strength components in addition to those given
in equation (\ref{kameq:nonzeroF}). In particular we choose $A_0^3(\varrho,x),\,A_3^1(\varrho,x)$
and $A_2^3(x)=x^1\,H_3^3$ yielding the additional components\footnote{Close to
the phase transition, it is consistent to drop the dependence of the field
$H^3_3$ on $\varrho$. Away from the phase transition the $\varrho$ dependence must
be included. From the boundary asymptotics it will be possible to extract the
magnetic field and the magnetization of the superconductor.}
\begin{eqnarray}
  \label{kameq:additionalF}
    &H_3^3=F^3_{12}=-F^3_{21}=\partial_1 A^3_2\, ,\\
    &F^1_{13}=-F^1_{31}=\partial_1 A^1_3\,,\\
    &F^2_{23}=-F^2_{32}=\frac{\gamma}{\sqrt{\lambda}}A^3_2A^1_3\,,\\
    &F^3_{1 0}=-F^3_{01}=\partial_1 A^3_0\,.
\end{eqnarray}
Recall that the radial AdS-direction is designated by
the indices $\varrho$ or $4$ synonymously.
Amending the DBI-action (\ref{kameq:DBI}) with the additional
components (\ref{kameq:additionalF}), we compute the determinant in analogy to
equation (\ref{kameq:DBI}). We then choose to expand the new action to second
order in $F$, i.e. we only consider terms being at most quadratic in the
fields. This procedure gives the truncated DBI action 
\begin{eqnarray}
   \label{kameq:DBIwithB}
   S_{\mathrm{DBI}}&=-T_{D7}N_f\int\!\mathrm{d}^8\xi\:\sqrt{-G}\Bigg[& 1+\frac{(2\pi\alpha')^2}{2} \Big(G^{00}G^{33} (F^2_{03})^2
    +G^{33}G^{44} (F^1_{\varrho 3})^2\nonumber\\ 
    &&+G^{00}G^{44}(F^3_{\varrho 0})^2
   +(G^{33})^2(F_{13}^1)^2+(G^{33})^2(F_{23}^2)^2\nonumber \\
    &&+G^{00}G^{33}(F_{10}^3)^2 +(G^{33})^2(F_{12}^3)^2\Big)+\cdots\Bigg]\,.
\end{eqnarray}
Respecting the symmetries and 
variable dependencies in our specific system, this can be written as
\begin{eqnarray}
   \label{kameq:DBIAwithB}
   S_{\mathrm{DBI}}&=-T_{D7}&N_f\int\!\mathrm{d}^8\xi\: \sqrt{-G}
   \Bigg[
   1+\frac{1}{2} \Big(G^{00}G^{33} (\partial_{\bar 1}\tilde A_0^3)^2+G^{33}G^{44} (\partial_\varrho \tilde A_3^1)^2 \\ \nonumber
   &&+ G^{00}G^{44}(\partial_{\varrho}\tilde A_0^3)^2
   +(G^{33})^2(\partial_{\bar 1} \tilde A_3^1)^2
   +(G^{33})^2 (\bar H_3^3)^2 \\ \nonumber
   &&+G^{00}G^{33}
   \frac{{\varrho_H}^4\gamma^2}{(2\pi\alpha')^2\lambda}
   (\tilde A_0^3\tilde A_3^1)^2  
   +(G^{33})^2\frac{{\varrho_H}^4\gamma^2}{(2\pi\alpha')^2\lambda}
    (\tilde A_3^1 \bar H_3^3)^2{\bar x}^2\Big)+\cdots\Bigg]\,,
\end{eqnarray}
with the convenient redefinitions
\begin{equation}
 \label{kameq:redefinitions}
 \tilde A=\frac{2\pi\alpha'}{\varrho_H} A\, ,\quad x=\varrho_H \bar x\, ,\quad
 \bar H_3^3=2\pi\alpha' H_3^3\,, \quad \varrho=\varrho_H\, \rho\,.
\end{equation}
Rescaling the $\bar x$-coordinate once more
\begin{equation}
\tilde x=\sqrt{\frac{\bar H_3^3 {\varrho_H}^2 \gamma}{2\pi\alpha'\sqrt{\lambda}}}\,\bar x\,,
\end{equation}
the equations of motion derived from the action (\ref{kameq:DBIAwithB}) take a simple form
\begin{eqnarray}
 \label{kameq:eomAB}
  0=&\partial_\rho^2\tilde A_0^3+\frac{\partial_\rho\left[\sqrt{-G} G^{00} G^{44}\right]}{\sqrt{-G} G^{00} G^{44}}
  \partial_\rho \tilde A_0^3+ \frac{\gamma \tilde H_3^3}{2\sqrt{2}\pi}\frac{G^{33}}{G^{44}}
  \partial_{\tilde x}\tilde A_0^3-\frac{\gamma^2}{2\pi^2}\frac{G^{33}}{G^{44}}
  \tilde A_0^3(\tilde A_3^1)^2\, ,\\ \nonumber
  0=&\partial_\rho^2\tilde A_3^1+\frac{\partial_\rho\left[\sqrt{-G} G^{33} G^{44}\right]}{\sqrt{-G} G^{33} G^{44}}
  \partial_\rho \tilde A_3^1 + \frac{\gamma \tilde H_3^3}{2\sqrt{2}\pi}\frac{G^{33}}{G^{44}}
  \left[\partial^2_{\tilde x}\tilde A_3^1-{\tilde x}^2\tilde A_3^1\right]-\frac{\gamma^2}{2\pi^2}\frac{G^{00}}{G^{44}}(\tilde A_0^3)^2\tilde A_3^1 \, .
\end{eqnarray}
Here all metric components are to be evaluated at $R=1$ and $\varrho\to\rho$.

We aim at decoupling and solving the system of partial differential
equations (\ref{kameq:eomAB}) by the product ansatz
\begin{equation}\label{kameq:prodAnsatz}
\tilde A_3^1(\rho,\tilde x)= v(\rho)\, u(\tilde x)\,.
\end{equation}
For this ansatz to work, we need to make two assumptions: First we assume 
that $\tilde A_0^3$ is constant in $\tilde x$. Second we assume that $A_3^1$
is small, which clearly is the case near the transition $T\to T_c$. Our second
assumption prevents $A_0^3$ from receiving a dependence on $\tilde x$ 
through its coupling to $A_3^1(\rho,\tilde x)$. These assumptions allow to
write the second equation in (\ref{kameq:eomAB}) as
\begin{eqnarray}
 \label{kameq:eomg}
  0=&\partial_\rho^2 v(\rho)+\frac{\partial_\rho\left[\sqrt{-G} G^{33} G^{44}\right]}{\sqrt{-G} G^{33} G^{44}}
  \partial_\rho v(\rho)-\frac{\gamma^2}{2\pi^2}\frac{G^{00}}{G^{44}}(\tilde A_0^3)^2 v(\rho)
  \nonumber\\   
  &+ \frac{\gamma \tilde H_3^3}{2\sqrt{2}\pi}\frac{G^{33}}{G^{44}}
  \,v(\rho)\,\frac{\partial^2_{\tilde x} u
  (\tilde x)-{\tilde x}^2 u(\tilde x)}{u
  (\tilde x)}\, .
\end{eqnarray}
All terms but the last one are independent of $\tilde x$, so the product 
ansatz (\ref{kameq:prodAnsatz}) is consistent only if 
\begin{equation}\label{kameq:eomf}
\frac{\partial^2_{\tilde x} u(\tilde x)-{\tilde x}^2 u(\tilde x)}{u(\tilde x)}=C\,,
\end{equation}
where $C$ is a constant. The differential equation (\ref{kameq:eomf}) has a
particular solution if $C=-(2n+1)\,, n\in \mathrm{N}$. The solutions for $u(\tilde x)$
are Hermite functions 
\begin{equation}
u_n(\tilde x)=\frac{e^{-\frac{|\tilde x|^2}{2}}}{\sqrt{n!2^n\sqrt{\pi}}}H_n(\tilde x)\,,\,\, 
 H_n(\tilde x)=(-1)^n e^{\frac{|\tilde x|^2}{2}}\frac{d^n}{dx^n} e^{-\frac{|\tilde x|^2}{2}},
\end{equation}
which have Gaussian decay at large $|\tilde x|\gg1$. 
Choosing the lowest solution with $n=0$ and $H_0=1$, which has no nodes, is most likely to give
the configuration with lowest energy content. 
So the system we need to solve is finally given by
\begin{eqnarray}
 \label{kameq:eomgA}
  0&=\partial_\rho^2\tilde A_0^3&+\frac{\partial_\rho\left[\sqrt{-G} G^{00} G^{44}\right]}{\sqrt{-G} G^{00} G^{44}}
  \partial_\rho \tilde A_0^3\frac{\gamma^2}{2\pi^2}\frac{G^{33}}{G^{44}}
 \tilde A_0^3(u_0(\tilde x)v(\rho))^2\, , \\ 
   0&=\partial_\rho^2 v&+\frac{\partial_\rho\left[\sqrt{-G} G^{33} G^{44}\right]}{\sqrt{-G} G^{33} G^{44}}
  \partial_\rho v-\frac{\gamma^2}{2\pi^2}\frac{G^{00}}{G^{44}}(\tilde A_0^3)^2 v
  -\frac{\gamma \tilde H_3^3}{2\sqrt{2}\pi}\frac{G^{33}}{G^{44}}\,v\, . 
\end{eqnarray}
Asymptotically near the horizon the fields take the form
\begin{eqnarray}\label{kameq:horAs}
  \,\tilde{A}_0^3 &=          &+a_2 (\rho-1)^2                        +\mathcal{O}((\rho-1)^{3})\, ,\\
  \,v &=b_0    &+\frac{b_0 H_3^3}{4} (\rho-1)^2+\mathcal{O}((\rho-1)^{-3})\,,
\end{eqnarray}
while at the boundary we obtain
\begin{eqnarray}\label{kameq:bdyAs}
  \,\tilde A_0^3 &=\tilde \mu &+\frac{\tilde d_0^3}{\rho^{2}}       +\mathcal{O}(\rho^{-4})\, ,\\
  \,v&=               &+\frac{\tilde d_3^1}{\rho^{2}}+\mathcal{O}(\rho^{-4})\, .
\end{eqnarray}
We succeed in finding numerical solutions $v(\rho)$ and
$A_0^3(\rho)$ to the set of equations (\ref{kameq:eomgA}) obeying the asymptotics
given by equations (\ref{kameq:horAs}) and (\ref{kameq:bdyAs}). These numerical
solutions are used to approach the phase transition from the
superconducting phase by increasing the magnetic field. We map out the line of critical
temperature-magnetic field pairs in figure \ref{kamfig:criticalBT}. In this way
we obtain a phase diagram displaying the Meissner effect. The critical line in
figure \ref{kamfig:criticalBT} separates the phase with and without
superconducting condensate $\tilde{d}_3^1$. 

We emphasize that this is a background  calculation involving no fluctuations.
Complementary to the procedure described above we also confirmed the phase diagram using  
the instability of the normal phase against fluctuations. Starting at large
magnetic field and vanishing condensate $\tilde d_3^1$, we determine for a
given magnetic field $H_3^3$ the temperature $T_c(H_3^3)$ at which the
fluctuation $a_1^3$ becomes unstable. That instability signals the
condensation process into the superconducting phase.

The presence of the coexistence phase below the critical line, where the system
is still superconducting despite the presence of an external magnetic field,
is the signal of the Meissner effect in the case of a global symmetry
considered here. If we now weakly gauged the flavor symmetry at the boundary,
the superconducting current $J_3^1$ would generate a magnetic field opposite
to the external field. Thus the phase observed is a necessary condition in the
case of a global symmetry for finding the standard Meissner effect when
gauging the symmetry.
%
\section{Interpretation \& Conclusion}
\label{kamsec:conclusion}
Our findings merge to a string theory picture of the \index{pairing mechanism}\emph{pairing mechanism} and 
the subsequent condensation process in section~\ref{kamsec:string-theory-pict}. 
Finally we summarize what we have learned about holographic p-wave super-somethings
and propose some future territories to be conquered. 
\subsection{String Theory Picture}
\label{kamsec:string-theory-pict}
We now develop a string theory interpretation, {i.\,e.\,} a geometrical
picture, of the formation of the superconducting/superfluid phase, for which the field theory is 
discussed in section \ref{kamsec:FTIdea}. We show that the system is stabilized by 
dynamically generating a non-zero vev of the current component
$J_3^1$ dual to the gauge field $A_3^1$ on the brane. 
Moreover, we find a geometrical picture of the pairing mechanism which
forms the condensate $\langle J_3^1\rangle$, the Cooper pairs.
\begin{figure}[b]
  \centering
  \includegraphics[width=0.5\linewidth]{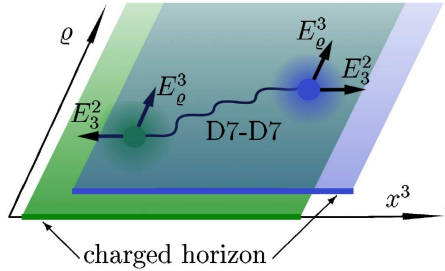}
  \caption{Sketch of our string setup: The figure shows the two
    coincident D7 branes stretched from the black hole horizon to the
    boundary as a
    green and a blue plane, respectively.  
Strings spanned from the horizon of the AdS
    black hole to the D$7$-branes  induce 
    a charge at the horizon
    \cite{kamErdmenger:2008yj,kamKobayashi:2006sb,kamKarch:2007br}. However, above a
    critical charge density, the strings charging the horizon recombine to
    D$7$-D$7$ strings. These D$7$-D$7$ strings are shown in the figure.
    Whereas the fundamental strings stretched between the horizon and the
    D$7$-brane are localized near the horizon, the D$7$-D$7$ strings propagate
    into the bulk balancing  the flavorelectric and 
    gravitational, {i.\,e.\,} tension forces (see text).  
    Thus these D$7$-D$7$ strings distribute
    the isospin charges along the AdS radial coordinate, leading to a
    stable configuration of reduced energy. This configuration of
    D$7$-D$7$ strings corresponds to a superconducting condensate.}
  \label{kamfig:stringpic}
\end{figure}
Let us first describe the unstable configuration in absence 
of the field $A^1_3$.
As known from~\cite{kamKarch:2007br,kamKobayashi:2006sb,kamErdmenger:2008yj}, the
non-zero field $A^3_0$ is induced by fundamental strings which are stretched
from the D$7$-brane to the horizon of the black hole. 
In the subsequent we call these strings `horizon strings'. 
Since the 
tension of these strings  would increase as they move to the boundary, they are
localized at the horizon, {i.\,e.\,} the horizon is effectively charged under
the isospin charge given by (\ref{kameq:7}). By increasing the horizon
string density, 
the isospin charge on the D7-brane at the horizon and therefore the energy
  of the system grows. In~\cite{kamErdmenger:2008yj}, the critical 
density was found beyond which this setup becomes unstable. In this case, the
strings would prefer to move towards the boundary due to the repulsive
force on their charged endpoints generated by the flavorelectric field
$E^3_\varrho=F^3_{0\varrho}=-\partial_\varrho A^3_0$. 

The setup is now stabilized by the new non-zero field $A^1_3$.
We can think of this field as being induced by D$7$-D$7$ strings moving in the $x^3$ direction.
This movement of the strings may be interpreted as a current in $x^3$
direction which induces  the magnetic field
$B^1_{3\varrho}=F^1_{3\varrho}=-\partial_\varrho A^1_3$. Moreover, the non-Abelian
interaction between the D$7$-D$7$ strings and the horizon strings induces a
flavorelectric field $E_3^2=F^2_{30}=\gamma/\sqrt{\lambda}A_0^3A_3^1$.

From the profile of the gauge fields and their conjugate momenta (see
figure~\ref{kamfig:profilesgaugefields}) we obtain the following: For $A_3^1=0$,
{i.\,e.\,} in the normal phase 
($T\ge T_c$), the isospin density $\tilde{d}_0^3$ is exclusively generated at 
the horizon by the horizon strings. 
This can also be understood by the profile of the conjugate
momenta $p_0^3$ (see figure~\ref{kamfig:profilesgaugefields} (bottom left)). We interpret
$p_0^3(\rho^*)$ as the isospin charge located between the horizon at $\rho=1$
and a fictitious boundary at $\rho=\rho^*$. 
In the normal phase, the momentum $p_0^3$ is
constant along the radial direction $\rho$ (see
figure~\ref{kamfig:profilesgaugefields}(bottom left)),
and therefore the isospin density is
exclusively generated at the horizon. In the superconducting phase where
$A_3^1\not=0$, {i.\,e.\,} $T<T_c$, the
momentum $p_0^3$ is not constant any more. Its value increases monotonically
towards the boundary and asymptotes to $\tilde{d}_0^3$ 
(see
figure~\ref{kamfig:profilesgaugefields}(bottom left)). 
Thus the isospin charge is also generated in the bulk
and not only at the horizon. This decreases the isospin charge 
at the horizon and stabilizes the system. 

Now we describe the string dynamics which distributes the
isospin charge into the bulk. Since the field $A_3^1$ induced by the D$7$-D$7$
strings is non-zero in the superconducting
phase, these strings must be responsible for
stabilizing this phase.  In the normal phase, there are only horizon
strings. In the superconducting phase, some of these strings
recombine to form D$7$-D$7$ strings which correspond to the non-zero gauge
field $A_3^1$ and carry isospin charge \footnote{Note that the
 D$7$-D$7$ strings are of the same order as the horizon
 strings, namely $N_f/N_c$, since they originate from the DBI action \cite{kamKarch:2007br}.}. 
There are two forces acting on the D$7$-D$7$ strings, the flavorelectric
force induced by the field $E_\varrho^3$ and the gravitational force between
the strings and the black hole. The flavorelectric force points to the
boundary while the gravitational force points to the horizon.  The
gravitational force is determined by the change in effective string 
tension, which contains the $\varrho$ dependent warp factor. The position of
the D$7$-D$7$ strings is determined by the equilibrium of these two forces.
Therefore the D$7$-D$7$ strings propagate from the horizon into the bulk and
distribute the isospin charge.

Since the D$7$-D$7$ strings induce the field $A_3^1$, they also generate the
density $\tilde{d}_3^1$ dual to the condensate $\langle J_3^1\rangle$, the
Cooper pairs. This density $\tilde{d}_3^1$
is proportional to the D$7$-D$7$ strings located in the bulk, 
in the same way as
the density $\tilde{d}_0^3$ counts the strings which carry isospin charge
\cite{kamKobayashi:2006sb}. This suggest that we can also interpret $p_3^1(\rho^*)$ as the number
of D$7$-D$7$ strings which are located between the horizon at $\rho=1$ and the
fictitious boundary at $\rho=\rho^*$. The momentum $p_3^1$ is always zero at the
horizon and increases monotonically in the bulk (see
figure~\ref{kamfig:profilesgaugefields} (bottom right)). Thus there are no D$7$-D$7$ strings
at the horizon, nevertheless most of them are located near the horizon.
\begin{svgraybox}
The double importance of the D$7$-D$7$ strings is given by the fact that they
are both responsible for stabilizing the superconducting
phase by lowering the isospin charge density at the horizon, 
as well as being the dual of the Cooper pairs since they break the
$U(1)_3$ symmetry. In QCD-language they correspond to quarks pairing up
to form charged vector mesons which condense subsequently. 
\end{svgraybox}

\subsection{Summary}
\label{kamsec:summary}
In conclusion we have derived from the top (string theory) down to the gravity theory a holographic
p-wave superconductor. Thus we were able to directly identify the degrees of freedom in the boundary field theory,
allowing us to translate geometric features directly into field theory features. In particular
we have found a string theory picture of the pairing mechanism. The Cooper pairs are 
modeled by strings spanned between the two flavor D$7$-branes corresponding to
quasi-particles in the vector bi-fundamental representation, i.\,e.\, vector mesons. The dual thermal field
theory is 3+1-dimensional
${\cal N}=2$ supersymmetric Yang-Mills theory with $SU(N_c)$ color and $SU(2)$ flavor symmetry
coupled to an ${\cal N}=4$ gauge multiplet. It shows a conductivity gap at low temperatures.
A pseudo-gap forms even above $T_c$. The onset of the Meissner-Ochsenfeld effect is visible
and in the conductivity spectrum we find massive quasi-particles even at vanishing quark mass.
Their masses are generated through a Higgs-like mechanism in the bulk. 

Note that our results
can also be interpreted using this very setup as a model for the quark gluon plasma as introduced
in section~\ref{kamsec:flavor}. In that case we have found a flavor superfluid phase. This should
not be confused with the color-superconducting phase theoretically found in QCD at 
high \emph{baryon density}.
%
\subsection{Outlook}
\label{kamsec:outlook}
Some directions for promising future investigation are the study of critical exponents near 
the phase transition. Transport coefficients such as the speeds of second, fourth and other sounds
can be determined. For some of these it will be neccessary or convenient to use the backreacted 
solutions for this D3/D7 system given in~\cite{kamAmmon:2009xh}. 
Similar to earlier semi-classical drag computations it might be instructive to compute the 
drag on the various strings near the superconducting phase transition. This computation had been suggested
already in~\cite{kamAmmon:2008fc} and carried out for a scalar condensate superfluid in~\cite{Gubser:2009qf}.
It would be interesting to see what happens to the Fermi surface formed by the
background fermions when the superconducting/superfluid phase is entered. 
Adventurous spirits may also take this setup as a serious model for p-wave superconductors such
as the ruthenate compounds mentioned in the introduction. The geometric insights we gain from the 
dual gravity description can in principle be directly translated to precise field theory statements.
Again this pleasant fact is due to knowing the field theory degrees of freedom exactly by using the gauge/gravity
correspondence. A different setup in which vector mesons condense at finite isospin potential is the Sakai-Sugimoto model as 
shown in~\cite{kamAharony:2007uu}. It could be illuminating to study all aforementioned effects in
this model as well~(\cite{kamRebhan:2008ur} studies a possibly superconducting phase in the
Sakai-Sugimoto model but considers a pion condensate at finite baryon density).

\begin{acknowledgement}
I thank Johanna Erdmenger for discussions as well as kind support, and Martin Ammon, 
Patrick Kerner, Felix Rust for fruitful collaboration. I thank Ronny Thomale for
instructive discussions, and especially Christopher Herzog and Andy O'Bannon for very 
helpful comments on these notes. This work is supported in part by the 
\emph{"Deutsche Forschungsgemeinschaft"~(DFG)}. Last but certainly not least, I thank the organizers and
participants of the \emph{Fifth Aegean Summer School "From Gravity to Thermal Gauge Theory: 
The AdS/CFT Correspondence"}.
\end{acknowledgement}
%

\providecommand{\href}[2]{#2}\begingroup\raggedright\endgroup

\end{document}